\documentclass[a4paper, 11pt]{article}

\usepackage{natbib}
\bibpunct{(}{)}{;}{a}{}{,}
\usepackage{setspace}
\usepackage{graphicx}
\usepackage{amsmath, amsthm, amssymb, amsfonts}
\usepackage{url}
\usepackage{multirow}
\usepackage{geometry}
\usepackage{epsfig}
\usepackage{psfrag}

\makeatletter
\def\url@leostyle{%
  \@ifundefined{selectfont}{\def\UrlFont{\sf}}{\def\UrlFont{\small\ttfamily}}}
\makeatother
\theoremstyle{definition}
\newtheorem{thm}{Theorem}
\theoremstyle{definition}

\theoremstyle{definition}
\newtheorem{lem}{Lemma}
\theoremstyle{definition}
\newtheorem{prop}{Proposition}
\theoremstyle{definition}

\theoremstyle{remark}

\theoremstyle{definition}
\newtheorem{defn}{Definition}

\newcommand{\sumi}{\sum_{i=-\infty}^{-1}}
\newcommand{\sumk}{\sum_{k=-\infty}^{\infty}}
\newcommand{\sumse}{\sum_{t=s}^e}

\newcommand{\ep}{\epsilon}
\newcommand{\vep}{\varepsilon}
\newcommand{\lam}{\lambda}
\newcommand{\sig}{\sigma}
\newcommand{\E}{\mathbb{E}}
\newcommand{\corr}{\mbox{cor}}
\newcommand{\cov}{\mbox{cov}}
\newcommand{\var}{\mbox{var}}
\newcommand{\tr}{\mbox{tr}}
\newcommand{\p}{\mathbb{P}}
\newcommand{\cN}{\mathcal{N}}
\newcommand{\cU}{\mathcal{U}}

\newcommand{\xtt}{X_{t, T}}
\newcommand{\bxtt}{\bX_{t, T}}
\newcommand{\xjtt}{X^{(j)}_{t, T}}
\newcommand{\ijitt}{I^{(j)}_{i, t, T}}
\newcommand{\zjstt}{Z^{(j)2}_{t, T}}
\newcommand{\ytt}{Y_{t, T}}
\newcommand{\yktt}{Y^{(k)}_{t, T}}
\newcommand{\tytt}{\tilde{Y}_{t, T}}
\newcommand{\ztt}{Z_{t, T}}
\newcommand{\tztt}{\tilde{Z}_{t, T}}
\newcommand{\zstt}{\ztt^2}
\newcommand{\zktt}{Z^{(k)}_{t, T}}
\newcommand{\zkstt}{Z^{(k)2}_{t, T}}
\newcommand{\sigtt}{\sig_{t, T}}
\newcommand{\sigt}{\sig(t/T)}
\newcommand{\sigktt}{\sig^{(k)}_{t, T}}
\newcommand{\sigkt}{\sig^{(k)}(t/T)}

\newcommand{\ty}{\tilde{y}}
\newcommand{\bA}{\mathbf{A}}
\newcommand{\bI}{\mathbf{I}}
\newcommand{\bV}{\mathbf{V}}
\newcommand{\bW}{\mathbf{W}}
\newcommand{\bX}{\mathbf{X}}
\newcommand{\bps}{\boldsymbol{\psi}}
\newcommand{\bxi}{\boldsymbol{\xi}}
\newcommand{\bep}{\boldsymbol{\epsilon}}
\newcommand{\bvep}{\boldsymbol{\varepsilon}}
\newcommand{\bSig}{\boldsymbol{\Sigma}}
\newcommand{\bet}{\boldsymbol{\eta}}
\newcommand{\bzero}{\mathbf{0}}

\newcommand{\cA}{\mathcal{A}}
\newcommand{\bbB}{\mathbb{B}}
\newcommand{\cB}{\mathcal{B}}
\newcommand{\cD}{\mathcal{D}}
\newcommand{\cI}{\mathcal{I}}
\newcommand{\cK}{\mathcal{K}}
\newcommand{\cY}{\mathcal{Y}}
\newcommand{\bbI}{\mathbb{I}}
\newcommand{\bbJ}{\mathbb{J}}
\newcommand{\bbS}{\mathbb{S}}
\newcommand{\bbY}{\mathbb{Y}}
\newcommand{\tbY}{\tilde{\mathbb{Y}}}

\newcommand{\is}{I^*}
\newcommand{\heta}{\wh{\eta}}
\newcommand{\teta}{\tilde{\eta}}
\newcommand{\hnu}{\wh{\nu}}
\newcommand{\thr}{\pi_T}
\newcommand{\delt}{\delta_T}
\newcommand{\ept}{\ep_T}

\def\wh{\widehat}

\title{
Multiple change-point detection for high-dimensional time series via Sparsified Binary Segmentation
}

\author{
Haeran Cho
\thanks{School of Mathematics, University of Bristol. %\protect \ E-mail: {\tt haeran.cho@bristol.ac.uk}}
}
and Piotr Fryzlewicz
\thanks{Department of Statistics, London School of Economics. %\protect \ E-mail: {\tt  p.fryzlewicz@lse.ac.uk}}
}
}

\date{}

\begin{document}

\maketitle

\begin{abstract}
Time series segmentation, a.k.a. multiple change-point
detection, is a well-established problem. However, few solutions
are designed specifically for high-dimensional situations.
In this paper, our interest is in segmenting the second-order structure
of a high-dimensional time series. In a generic step of a binary
segmentation algorithm for multivariate time series, one natural
solution is to combine CUSUM statistics obtained from local periodograms and
cross-periodograms of the components of the input time series.
However, the standard ``maximum'' and ``average'' methods for
doing so often fail in high dimensions when, for example, the change-points
are sparse across the panel or the CUSUM statistics are spuriously large.

In this paper, we propose the Sparsified Binary Segmentation (SBS) algorithm
which aggregates the CUSUM statistics by adding only those that pass a
certain threshold. This ``sparsifying'' step reduces the impact of irrelevant,
noisy contributions, which is particularly beneficial in high dimensions.

In order to show the consistency of SBS, we introduce the multivariate Locally
Stationary Wavelet model for time series, which is a separate contribution of
this work.
\end{abstract}

\section{Introduction}
\label{sec:intro}

Detecting multiple change-points in univariate time series has been widely discussed in various contexts,
see \citet{inclan1994}, \citet{chen1997}, \citet{lavielle2000}, \citet{ombao2001},
\citet{davis2006} and \citet{davis2008} for some recent approaches.
In this article,
we use the term ``multiple change-point detection'' interchangeably with ``segmentation''.
By contrast, segmentation of the second-order structure of multivariate time series,
especially those of high dimensionality, is yet to receive much attention despite the fact that
multivariate time series observed in practical problems often
appear second-order nonstationary.
For example, in financial time series, large panels of asset returns routinely display such nonstationarities
(see e.g. \citet{flq11} for a comprehensive review of challenges of high-dimensionality
in finance and economics). Another example can be found in neuroscience,
where electroencephalograms (EEG) recorded at multiple channels exhibit
nonstationarity and high correlations as well as being massive in volume \citep{ombao2005}.
\citet{vert2010} describe other interesting examples of multivariate, nonstationary time
series in many other fields, such as signal processing, biology and medicine.

As arguably one of the simplest forms of departure from stationarity,
we consider a class of piecewise stationary, multivariate (possibly high-dimensional) time series
with a time-varying second-order structure,
where the autocovariance and cross-covariance functions are asymptotically piecewise constant and hence
the time series is approximately stationary between change-points in these functions.

We first list some existing approaches to
 the problem of multiple change-point detection in multivariate (not necessarily high-dimensional) time series.
\citet{ombao2005} employed the SLEX (smooth localized complex exponentials) basis for time series segmentation,
originally proposed by \citet{ombao2002}. The choice of SLEX basis leads to the segmentation of the time series,
achieved via complexity-penalized optimization.
\citet{lavielle2006} introduced a procedure based on penalized Gaussian log-likelihood as a cost function,
where the estimator was computed via dynamic programming. The performance of the method was tested on
bivariate examples.
\citet{vert2010} proposed a method for approximating multiple signals, with independent noise, via
piecewise constant functions, where the change-point detection problem was re-formulated as a
penalized regression problem and
solved by the group Lasso \citep{yuan2006}. Note that in \citet{cho2011}, we argued that the $\mathit{l}_1$-penalty
was sub-optimal for change-point detection.

CUSUM-type statistics have been widely used in time series segmentation.
In the context of multivariate time series segmentation,
\citet{groen2009} studied the average and the maximum of $d$ CUSUM statistics, each obtained from one
component of a $d$-dimensional time series, and compared their theoretical properties as well as finite
sample performance.
The average test statistic was also adopted in \citet{horvath2012} for detecting a single change
in the mean of a panel data model,
and both papers allowed the dimensionality to increase under the constraint $d^2/T \to 0$,
where $T$ denoted the sample size.
In \citet{aue2009}, a CUSUM statistic was proposed for detecting and locating a single change-point
in the covariance structure of multivariate time series, where its extension to the detection of multiple change-points
via binary segmentation was discussed heuristically.

In this paper, we propose a CUSUM-based binary segmentation algorithm,
termed ``Sparsified Binary Segmentation'' (SBS), for identifying multiple change-points
in the second-order structure of a multivariate (possibly high-dimensional) time series.
The input to the SBS algorithm is $\{Y_{t,T}^{(k)}, \, k = 1, \ldots, d\}$,
a $d$-dimensional sequence of localized periodograms and cross-periodograms computed on the original multivariate time series,
where the dimensionality $d$ is allowed to diverge with the number of observations $T$ at a certain rate.

A key ingredient of the SBS algorithm is a ``sparsifying'' step,
where, instead of blindly aggregating all the information about the change-points from the $d$ sequences
$Y_{t,T}^{(k)}$,
we apply a threshold to the individual CUSUM statistics computed on each $Y_{t,T}^{(k)}$,
and only those temporal fragments of the CUSUMs that survive after the thresholding are aggregated
to have any contribution in detecting and locating the change-points.
In this manner, we reduce the impact of those sequences that do not contain any change-points
so that the procedure is less affected by them,
which can be particularly beneficial in a high-dimensional context.
Therefore, we can expect improved performance in comparison to methods without
a similar dimension-reduction step, and this point is explained in more detail in Section
\ref{sec:malg}. Further, due to the aggregation of the CUSUM statistics, the algorithm
automatically identifies common change-points, rather than estimating single change-points
at different locations in different components of the time series, which removes the
need for post-processing across the $d$-dimensional sequence.
This latter characteristic is particularly attractive in a high-dimensional situation.

As well as formulating the complete SBS algorithm, we show its
consistency for the number and the locations of the change-points.
One theoretical contribution of this work is that our rates of convergence
of the location estimators improve on those previously obtained for
binary segmentation for univariate time series \citep{cho2012}
and are near-optimal in the case of the change-points being
separated by time intervals of length $\asymp T$,
where $a_T \asymp b_T$ if $a_T^{-1}b_T \to C$ as $T \to \infty$ for some constant $C$.
This was achieved by adapting, to the high-dimensional time series context, the
proof techniques from \citet{piotr2013} for the univariate signal plus i.i.d. Gaussian noise model.

As a theoretical setting for deriving the consistency results, we introduce the multivariate Locally Stationary Wavelet (LSW) model for time series.
This, we believe, is a separate contribution of the current work, and provides a
multivariate extension of the univariate LSW model of \citet{nason2000}
and of the bivariate LSW model of \citet{jean2010}.

The rest of the paper is organized as follows.
In Section \ref{sec:seg}, we introduce the SBS algorithm for segmenting a possibly large number of
multiplicative sequences. In Section \ref{sec:link}, we introduce a class of piecewise stationary,
multivariate time series and discuss the specifics of applying the SBS from Section \ref{sec:seg} to detect change-points
in its second-order structure (the version of SBS specifically applicable to multivariate time series
is labeled SBS-MVTS in the paper). Section \ref{sec:sim} illustrates the performance of the proposed methodology
on a set of simulated examples, and Section \ref{sec:real} applies it to the multivariate series of S\&P 500
components, observed daily between 2007 and 2011. The proofs are in the Appendix.

\section{The SBS algorithm in a generic setting}
\label{sec:seg}

In this section, we outline the SBS algorithm for change-point detection
in a panel of multiplicative sequences, which may share common change-points in their expectations.
We later consider a piecewise stationary, multivariate time series model and use it to derive a set of statistics,
which contain information about the change-points in its second-order structure. Those statistics
are shown to follow the multiplicative model considered so that SBS can be applied to them. This will enable us to
segment the original time series using the SBS methodology.

The multiplicative model in question is
\begin{eqnarray}
\yktt=\sigkt\cdot\zkstt, \quad t=0, \ldots, T-1; \ k=1, \ldots, d,
\label{generic:one}
\end{eqnarray}
where $\zktt$ is a sequence of (possibly) autocorrelated and nonstationary standard normal variables such that $\E\yktt=\sigkt$,
which implies that each $\yktt$ is a scaled $\chi^2_1$ variable. Extensions to some other
distributions are possible but technically involved and we do not pursue them here.
Each $\sigkt$ is a piecewise constant function, and we aim to detect any change-points
in $\sigkt$ for $k=1, \ldots, d$. It is assumed that there are $N$ change-points
$0<\eta_1<\eta_2<\ldots<\eta_N<T-1$ possibly shared by the $d$ functions $\sigkt$,
in the sense that for each $\eta_q$, there exists one or more $\sigkt$ satisfying
$\sig^{(k)}(\eta_q/T) \ne \sig^{(k)}((\eta_q+1)/T)$. We impose the following conditions on $\eta_q, \ q=1, \ldots, N$.
\begin{itemize}
\item[(A1)]
\begin{itemize}
\item[(i)] The distance between any two adjacent change-points is bounded from below by
$\delt \asymp T^\Theta$ for $\Theta \in (3/4, 1]$.
\item[(ii)] The spacings between any three consecutive change-points are not too ``unbalanced'' in
the sense that they satisfy
\begin{eqnarray}
\max\left(\frac{\eta_q-\eta_{q-1}+1}{\eta_{q+1}-\eta_{q-1}+1},
\frac{\eta_{q+1}-\eta_q}{\eta_{q+1}-\eta_{q-1}+1}\right) \le c_*,
\label{assum:eq}
\end{eqnarray}
where $c_*$ is a constant satisfying $c_* \in [1/2, 1)$.
\end{itemize}
\end{itemize}
Note that (A1.i) determines the upper bound on the total number of change-points,
which is allowed to diverge with $T$ as long as $\Theta<1$, and is unknown by the user.
\citet{cho2012} proposed a change-point detection method for a \emph{single} sequence $\ytt$ following model (\ref{generic:one}).
The main ingredient of the method proposed in that work was a binary segmentation algorithm
which simultaneously located and tested for change-points in a recursive manner.
Below we provide a sketch of that algorithm, which is referred to as Univariate Binary Segmentation (UBS)
throughout the present paper.

Firstly, the likely position of a change-point in the interval $[0, T-1]$ is located
as the point where the following CUSUM-type statistic is maximized over $t$;
\begin{eqnarray}
\cY_{0, t, T-1}
= \cY_{0, t, T-1}(Y_{u,T})
=\left(\frac{1}{T}\sum_{u=0}^{T-1} Y_{u, T}\right)^{-1} \cdot
\left\vert\sqrt{\frac{T-t}{T \cdot t}} \sum_{u=0}^{t-1} Y_{u, T} -
\sqrt{\frac{t}{T\cdot(T-t)}}\sum_{u=t}^{T-1} Y_{u, T}\right\vert.
\label{unbalhaar}
\end{eqnarray}
A discussion of the properties of $\cY_{0, t, T-1}$ can be found in \citet{cho2012};
we only remark here that the first term of the product in (\ref{unbalhaar})
is a normalizing term essential in multiplicative settings, which makes our results independent
of the level of $\sigkt$ in (\ref{generic:one}). Next,
for $b = \arg\max_t \cY_{0, t, T-1}$, if $\cY_{0, b, T-1} < \thr$ with a suitably chosen threshold $\thr$,
then we stop; otherwise we add $b$ to the set of estimated change-points and continue recursively in
the same manner to the left and to the right of $b$. Details of the UBS algorithm and the theoretical
result on its consistency for the number and the locations of the change-points can be found in
the above work.

\subsection{Binary segmentation for high-dimensional data}
\label{sec:malg}

In this section, we extend the UBS algorithm to one which is applicable to a panel of multiplicative sequences
(\ref{generic:one}) even if its dimensionality $d$ diverges as $T \to \infty$. The resulting
SBS algorithm contains a crucial ``sparsifying'' step as detailed below.

We firstly note that in the multivariate case $d > 1$, we could proceed by applying the UBS algorithm
to each sequence $\yktt$ separately,
and then pruning the estimated change-points by identifying those corresponding to each
true change-point. However, it is conceivable that such pruning may not be straightforward,
particularly in high dimensions. We propose to circumvent this difficulty by segmenting the $d$
sequences $\yktt$ at the same time by examining the CUSUM statistics $\cY_{0, t, T-1}(Y_{u,T}^{(k)})\equiv\cY^{(k)}_{0, t, T-1}$
in (\ref{unbalhaar}) simultaneously over $k$, rather than separately for each $k$.

A number of ways of aggregating information from multiple CUSUM statistics have been proposed in the
literature.
\citet{groen2009} discussed two popular methods: the point-wise average, and the point-wise maximum.
Specifically, using our notation, they are respectively defined as
\begin{eqnarray}
\ty^{\mbox{\scriptsize avg}}_t = \frac{1}{d} \sum_{k=1}^d \cY^{(k)}_{0, t, T-1}, \qquad
\ty^{\mbox{\scriptsize max}}_t = \max_{1 \le k \le d} \cY^{(k)}_{0, t, T-1}.
\label{eq:groen}
\end{eqnarray}
To determine whether $b = \arg\max \ty^{\mbox{\scriptsize avg}}_t$ ($\ty^{\mbox{\scriptsize max}}_t$)
is regarded as an estimated change-point, $\ty^{\mbox{\scriptsize avg}}_b$ ($\ty^{\mbox{\scriptsize max}}_b$)
needs to be compared against a threshold which takes into account the aggregation step.

In the SBS algorithm, we propose another way of simultaneously considering multiple CUSUM statistics, which integrates a
thresholding step that enables us to bypass some difficulties in dealing with high-dimensional data which we describe later on.
For each $k$, the CUSUM statistic $\cY^{(k)}_{0, t, T-1}$ is compared with a threshold, say $\thr$
(to be specified later in Section \ref{sec:link}), and
only the contributions from the time intervals where $\cY^{(k)}_{0, t, T-1} > \thr$ are taken into account
in detecting and locating a change-point. Thus $\ty^{\mbox{\scriptsize thr}}_t$, the main statistic
of interest in the SBS algorithm, is defined as
\begin{eqnarray}
\ty^{\mbox{\scriptsize thr}}_t = \sum_{k=1}^d \cY^{(k)}_{0, t, T-1} \cdot \bbI\left(\cY^{(k)}_{0, t, T-1} > \thr\right),
\label{eq:thr}
\end{eqnarray}
where $\bbI(\cdot)$ is an indicator function returning $\bbI(\cA)=1$ if the event $\cA$ is true and $\bbI(\cA)=0$ otherwise.
In this manner, $\ty^{\mbox{\scriptsize thr}}_t$ is non-zero only when
at least one of $\cY^{(k)}_{0, t, T-1}$ is greater than the threshold,
i.e. a change-point is detected in $\yktt$ for such $k$.
Therefore we can conclude that a change-point is detected in the $d$-dimensional multiplicative sequences and,
without applying any pruning, its location is estimated as $b=\arg\max_t\ty^{\mbox{\scriptsize thr}}_t$.

While the empirical study conducted in \citet{groen2009} shows the effectiveness of
both $\ty^{\mbox{\scriptsize avg}}_t$ and $\ty^{\mbox{\scriptsize max}}_t$ in detecting the presence of a single change-point,
there exist high-dimensional scenarios where these two estimators fail.
Below we provide examples of high-dimensional situations where
$\ty^{\mbox{\scriptsize thr}}_t$ exhibits better performance than the other two.

\begin{description}
\item[(A) Sparse change-points.] \hfill \\
We first independently generate two time series $X^{(k)}_t, k=1, 2$ as
\begin{eqnarray*}
X^{(1)}_{t, T} &=& aX^{(1)}_{t-1, T} + \ep^{(1)}_{t, T} \\
X^{(2)}_{t, T} &=& \left\{\begin{array}{ll}
0.95X^{(2)}_{t-1, T} + \ep^{(2)}_{t, T} & \mbox{for } 1 \le t \le \lfloor T/2 \rfloor, \\
0.3X^{(2)}_{t-1, T} + \ep^{(2)}_{t, T} & \mbox{for } \lfloor T/2 \rfloor +1 \le t \le T,
\end{array}\right.
\end{eqnarray*}
with $T=1024$.
The parameter $a$ is randomly generated from a uniform distribution $\cU(0.5, 0.99)$ and
$\ep^{(k)}_{t, T}$ are i.i.d. standard normal variables for $k=1, 2$.
We further produce the sequences $Y^{(1)}_{t, T}$ and $Y^{(2)}_{t, T}$ as
$Y^{(k)}_{t, T} = 2^{-1}(X^{(k)}_{t, T}-X^{(k)}_{t-1, T})^2, \ k=1, 2$,
such that $Y^{(1)}_{t, T}$ does not have any change in $\E Y^{(1)}_{t, T}$,
while $\E Y^{(2)}_{t, T}$ has one change-point at $t=\lfloor T/2 \rfloor$.
The rationale behind the choice of $Y^{(k)}_{t, T}$ as well as its relationship to the multiplicative model (\ref{generic:one})
are discussed in detail in Section \ref{sec:link}.
As can be seen from the top panel of Figure \ref{fig:thr:a}, all three of the corresponding statistics
$\ty^{\mbox{\scriptsize avg}}_t$, $\ty^{\mbox{\scriptsize max}}_t$ and $\ty^{\mbox{\scriptsize thr}}_t$ are able to correctly identify the location of the true change-point.

Now, consider the case with $d=100$ time series where the additional time series
$X^{(k)}_{t, T}, \ k=3, \ldots, d$ are independently generated as $X^{(1)}_{t, T}$
such that, overall, there is only one change-point
coming from $X^{(2)}_{t, T}$ in the entire panel.
Then, in obtaining the point-wise average of the $d$ CUSUM statistics
in $\ty^{\mbox{\scriptsize avg}}_t$, the $\cY^{(k)}_{0, t, T-1}$ for $k \ne 2$ corrupt the peak
that is achieved around $t = \lfloor T/2 \rfloor$ for $\cY^{(2)}_{0, t, T-1}$,
and hence the maximum of $\ty^{\mbox{\scriptsize avg}}_t$ is attained far from the true change-point.
On the other hand, both $\ty^{\mbox{\scriptsize thr}}_t$ and $\ty^{\mbox{\scriptsize max}}_t$
are successful in maintaining the peak achieved by $\cY^{(2)}_{0, t, T-1}$ by disregarding most or all
of the $\cY^{(k)}_{0, t, T-1}, \ k \ne 2$.

\begin{figure}[htbp]
\centering
\begin{tabular}{c}
\epsfig{file=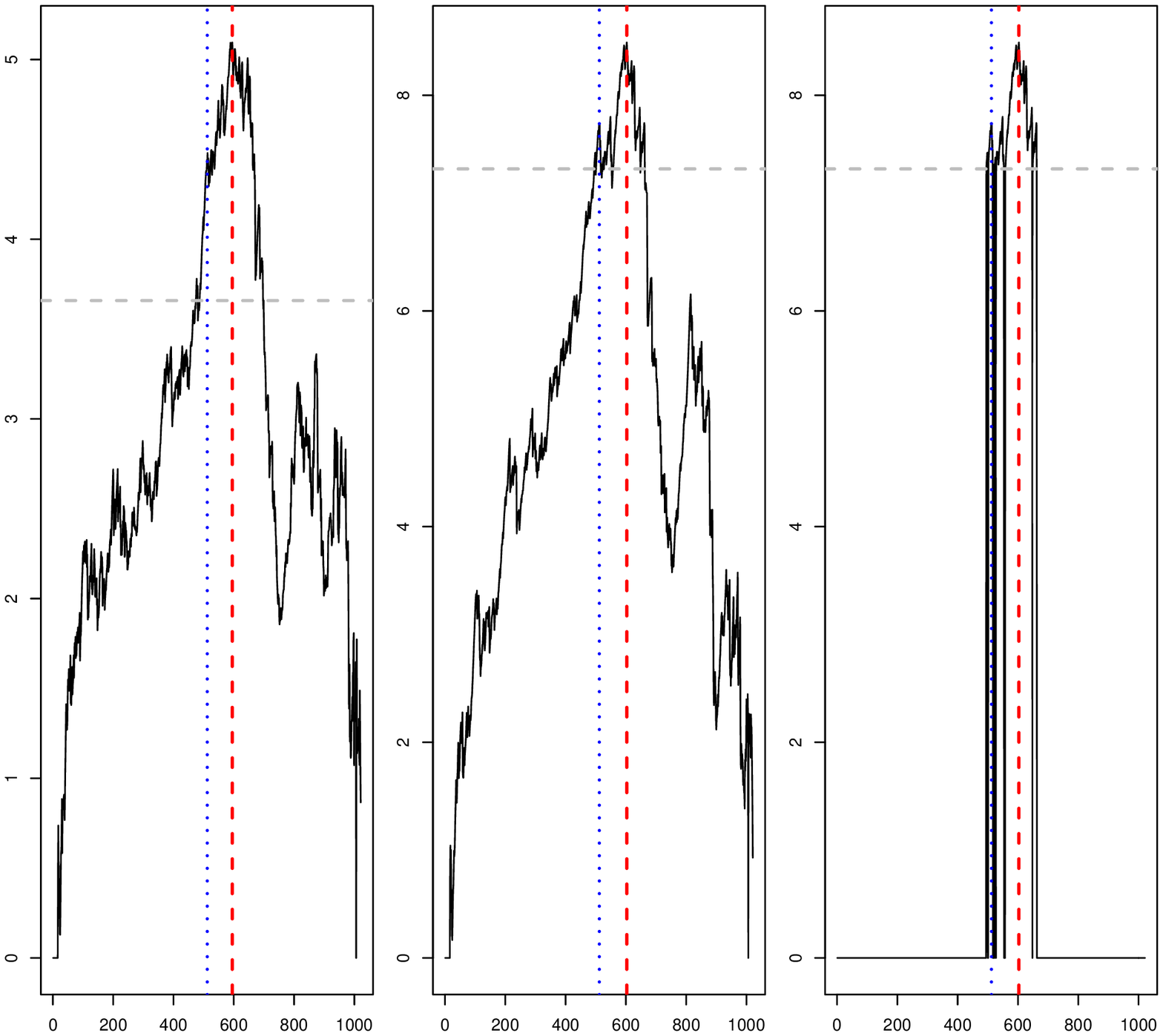, width=1\linewidth, height=2.25in} \\
\epsfig{file=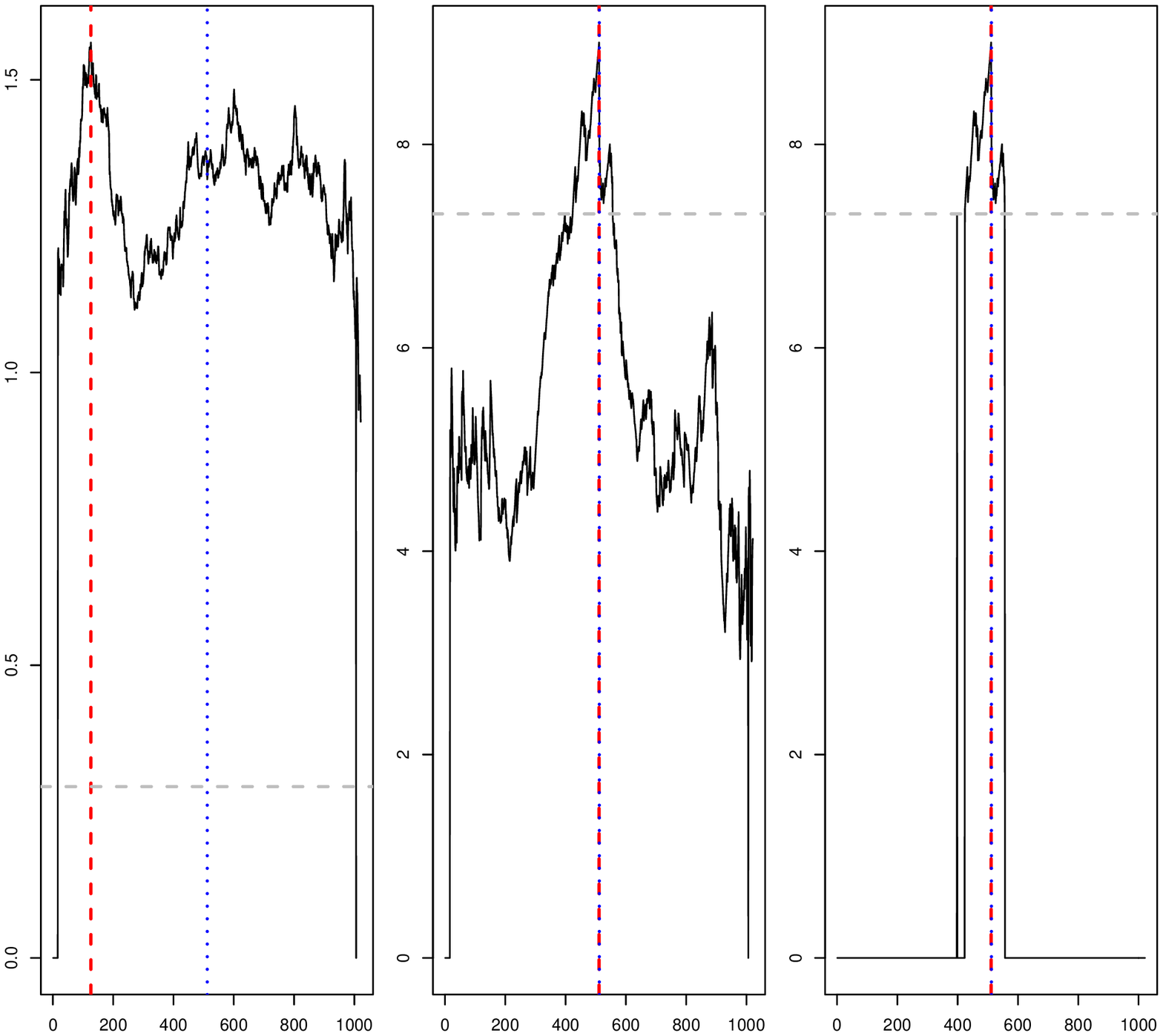, width=1\linewidth, height=2.25in}
\end{tabular}
\caption{
Top: $\ty^{\mbox{\scriptsize avg}}_t$ (left), $\ty^{\mbox{\scriptsize max}}_t$ (center), and $\ty^{\mbox{\scriptsize thr}}_t$ (right)
from model (A) in Section \ref{sec:malg} when $d=2$;
bottom: $d=100$;
the broken vertical lines: location of the maximum of each of these sequences;
the dotted vertical lines: location of the true change-point;
the broken horizontal lines: the threshold $\thr$.}
\label{fig:thr:a}
\end{figure}

\item[(B) Spuriously large CUSUM statistics.] \hfill \\
Again, we first independently generate $d=2$ time series $X^{(k)}_{t, T}, k=1, 2$,
 with $T=1024$, where $X^{(1)}_{t, T}$ is identical to that in (A), and
\begin{eqnarray*}
X^{(2)}_{t, T} &=& \left\{\begin{array}{ll}
0.3X^{(2)}_{t-1, T} + \ep^{(2)}_{t, T} & \mbox{for } 1 \le t \le 100, \\
-0.75X^{(2)}_{t-1, T} + \ep^{(2)}_{t, T} & \mbox{for } 101 \le t \le T.
\end{array}\right.
\end{eqnarray*}
$X^{(2)}_{t, T}$ is composed of two stationary segments, where the first segment is relatively short and
(weekly) positively autocorrelated, and the second one is long and negatively autocorrelated.
The negative autocorrelation in $X^{(2)}_{t, T}$
for $t \ge 101$ leads to $Y^{(2)}_{t, T}$ being highly autocorrelated,
which in turn results in spuriously large values of $\cY^{(2)}_{0, t, T-1}$ for $t \ge 101$ even when $t$ is far from the true change-point.
However, when $d=2$, all three statistics
$\ty^{\mbox{\scriptsize thr}}_t$, $\ty^{\mbox{\scriptsize max}}_t$ and $\ty^{\mbox{\scriptsize avg}}_t$
still manage to locate the true change-point around $t=100$, which is illustrated in
the top panel of Figure \ref{fig:thr:b}.

Now, let $d=100$ and independently generate 50 time series distributed
as $X^{(1)}_{t, T}$ and 50 as $X^{(2)}_{t, T}$
such that the change-point is not sparse across the panel.
Since there are $d/2 = 50$ sequences $Y^{(k)}_{t, T}$ for which
the CUSUM statistics $\cY^{(k)}_{0, t, T-1}$ can take spuriously large
values anywhere over $t\in[101, T]$, the statistic
$\ty^{\mbox{\scriptsize max}}_t$ becomes corrupted and is no longer able
to identify the true change-point.

On the other hand, $\ty^{\mbox{\scriptsize thr}}_t$ not only disregards
the contribution from the segments containing no change-points,
but also aggregates the contribution from those containing the
change-point, and therefore is able to identify the change-point very clearly.
In this example, the aggregation effect also causes $\ty^{\mbox{\scriptsize avg}}_t$
to work well.

\begin{figure}[htbp]
\centering
\begin{tabular}{c}
\epsfig{file=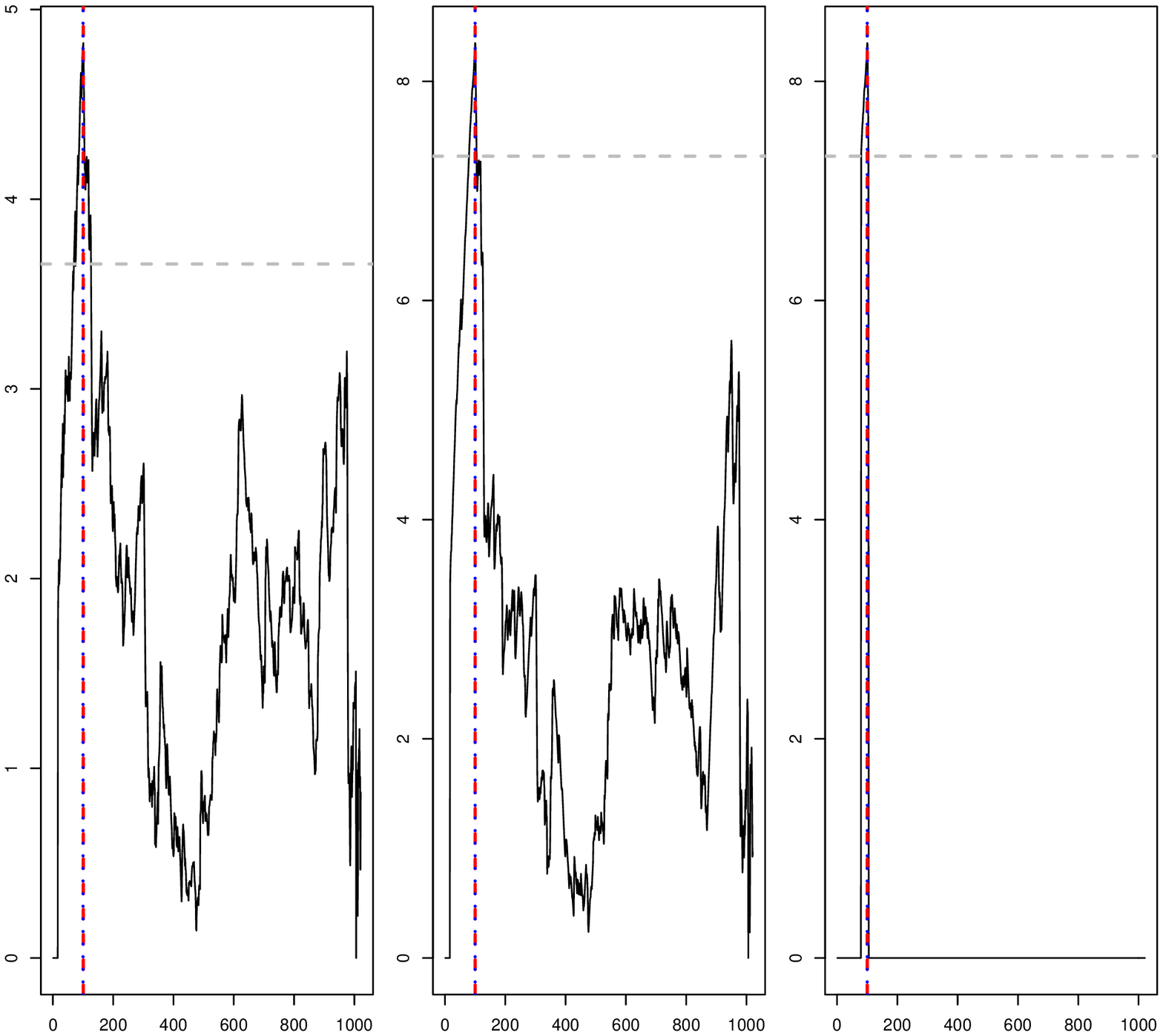, width=1\linewidth, height=2.25in} \\
\epsfig{file=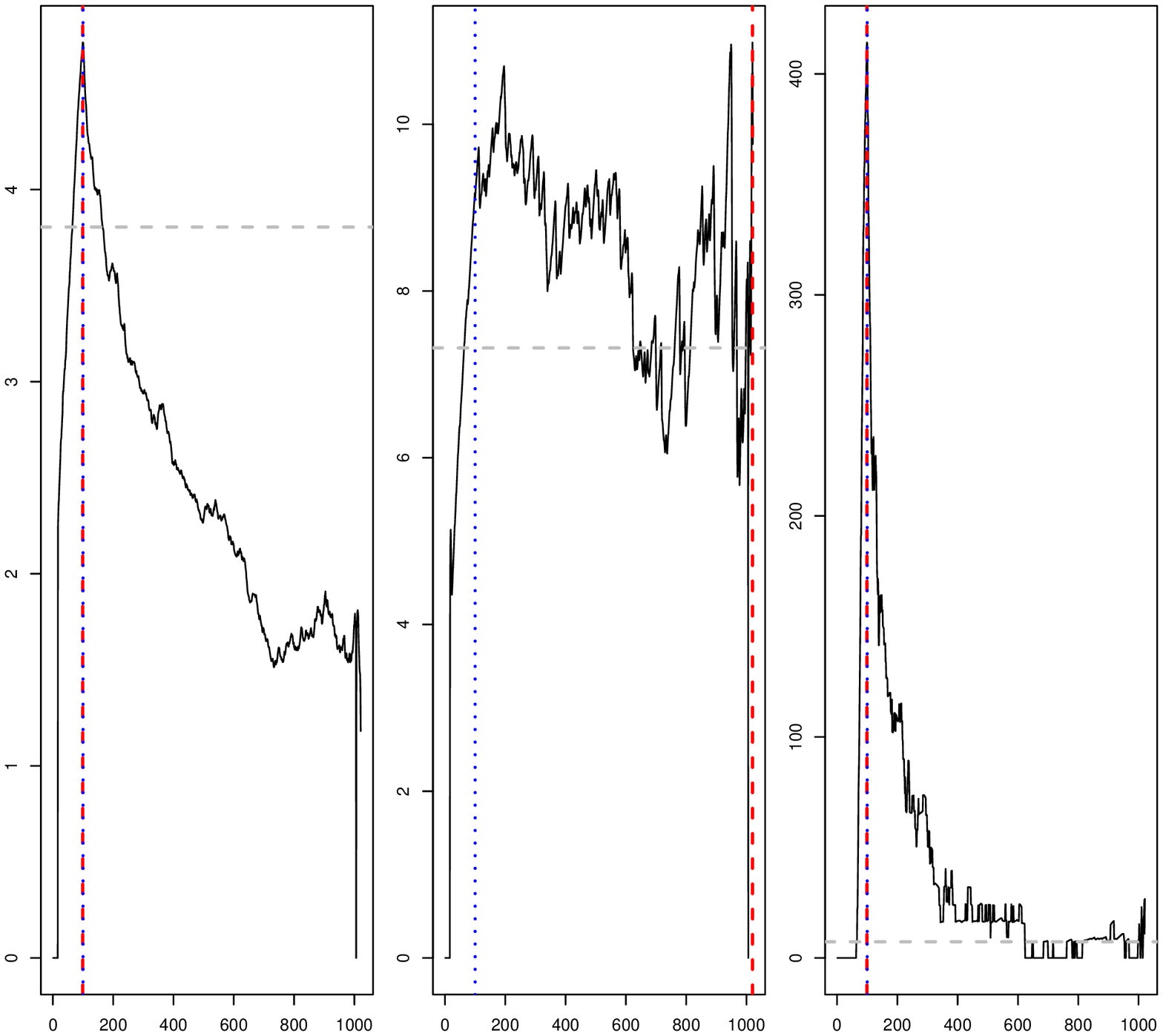, width=1\linewidth, height=2.25in}
\end{tabular}
\caption{
Top: $\ty^{\mbox{\scriptsize avg}}_t$ (left), $\ty^{\mbox{\scriptsize max}}_t$ (center), and $\ty^{\mbox{\scriptsize thr}}_t$ (right)
from (B) in Section \ref{sec:malg} when $d=2$;
bottom: $d=100$;
the broken vertical lines: location of the maximum of each of these sequences;
the dotted vertical lines: location of the true change-point;
the broken horizontal lines: the threshold $\thr$.}
\label{fig:thr:b}
\end{figure}

\end{description}

To summarize, $\ty^{\mbox{\scriptsize thr}}_t$ is shown to be better at dealing with
some difficulties arising from the high-dimensionality of the data than either
$\ty^{\mbox{\scriptsize avg}}_t$ or $\ty^{\mbox{\scriptsize max}}_t$ in these two
examples. In addition, the superior performance of $\ty^{\mbox{\scriptsize thr}}_t$
is attributed to different features of the sparsifying step in the two cases.

Motivated by the above discussion, we now introduce our SBS algorithm for segmenting
$d$-dimensional series below.
We use $j$ to denote the level index (indicating the progression of the segmentation procedure)
and $l$ to denote the location index of the node at each level.

\begin{description}
\item[SBS algorithm]
\item[Step 0]
Start with $(j, l)=(1, 1)$, setting $s_{1, 1}=0$, $e_{1, 1}=T-1$ and $n_{1, 1}=e_{1, 1}-s_{1, 1}+1$.

\item[Step 1]
Compute the CUSUM statistics $\cY^{(k)}_{s_{j, l}, t, e_{j, l}}$ as in (\ref{unbalhaar}) for all $k=1, \ldots, d$
over $t\in(s_{j, l}, e_{j, l})$, and obtain $\ty^{\mbox{\scriptsize thr}}_t$ as
\begin{eqnarray}
\ty^{\mbox{\scriptsize thr}}_t = \sum_{k=1}^d \cY^{(k)}_{s_{j, l}, t, e_{j, l}} \cdot \bbI\left(\cY^{(k)}_{s_{j, l}, t, e_{j, l}} > \thr\right),
\nonumber
\end{eqnarray}
with a threshold $\thr$.

\item[Step 2] \hfill
\begin{description}
\item[Step 2.1] If $\ty^{\mbox{\scriptsize thr}}_t=0$ for all $t \in (s_{j, l}, e_{j, l})$, stop the algorithm for the interval $[s_{j, l}, e_{j, l}]$.
\item[Step 2.2] If not, find $t$ that maximizes the corresponding $\ty^{\mbox{\scriptsize thr}}_t$ while satisfying
\begin{eqnarray}
\max \left( \frac{t-s_{j, l}+1}{n_{j, l}}, \frac{e_{j, l}-t}{n_{j, l}}\right) \le c_*,
\label{uh:cond}
\end{eqnarray}
where $c_*$ is identical to the one in (A1).
\item[Step 2.3] If there exists any $u\in[t-\Delta_T, t+\Delta_T]$ for which $\ty^{\mbox{\scriptsize thr}}_u=0$,
go back to Step 2.2 and find $t$ attaining the next largest $\ty^{\mbox{\scriptsize thr}}_t$ while satisfying (\ref{uh:cond}).
Repeat the above until a $t$ is found that satisfies $\ty^{\mbox{\scriptsize thr}}_u > 0$ for all $u\in[t-\Delta_T, t+\Delta_T]$,
set such $t$ as $b_{j, l}$ and proceed to Step 3.
If such $t$ does not exist, stop the algorithm for the interval $[s_{j, l}, e_{j, l}]$.
\end{description}

\item[Step 3]
Set $b_{j, l}$ as an estimated change-point
and divide the interval $[s_{j, l}, e_{j, l}]$ into two subintervals
$(s_{j+1, 2l-1}, e_{j+1, 2l-1}) \leftarrow (s_{j, l}, b_{j, l})$ and
$(s_{j+1, 2l}, e_{j+1, 2l}) \leftarrow (b_{j, l}+1, e_{j, l})$.
Update the level $j$ as $j \leftarrow j+1$ and go to Step 1.
\end{description}
Condition (\ref{uh:cond}) is imposed to prevent the algorithm from
detecting a change-point that is too close to previously detected ones;
note that in (A1), a similar condition is imposed on the locations of the true change-points.

As seen in Section \ref{sec:malg} with two motivating examples,
the performance of a change-point detection method for high-dimensional time series
depends on many factors besides the underlying dimension,
and we cannot set $\thr$ to uniformly increase or decrease with $d$.
Instead, to handle the false alarms in multiple testing procedure,
the threshold $\thr$ is derived such that
on any segment $[s, e]$ containing previously undetected true change-points for at least one $k=1, \ldots, d$,
the test statistic $\max_{t\in(s, e)}\cY^{(k)}_{s, t, e}$ exceeds $\thr$ with probability
converging to one for all such $k$, while $\cY^{(k)}_{s, t, e} < \thr$ for the remaining $k$'s,
as long as $d$ satisfies (A4).

Also, as the CUSUM statistic $\cY^{(k)}_{s_{j, l}, t, e_{j, l}}$ is expected to increase and then decrease
smoothly around true change-points without discontinuities,
Step 2.3 ensures that the algorithm disregards any spurious spikes in $\cY^{(k)}_{s_{j, l}, t, e_{j, l}}$.
Section \ref{sec:thr} provides a detailed discussion on the
practical selection of the parameters of SBS, including
$\thr$ and $\Delta_T$.
Steps 2.1 and 2.3 provide a stopping rule for the algorithm on those intervals $[s_{j, l}, e_{j, l}]$
where either no CUSUM statistic $\cY^{(k)}_{s_{j, l}, t, e_{j, l}}$ exceeds $\thr$ (Step 2.1),
or the exceedance is judged to be spurious (Step 2.3).

As an aside, we note that the mechanics of the SBS algorithm can be applicable in more general situations too,
beyond the particular model (\ref{generic:one}).

\subsection{Consistency of the SBS algorithm}
\label{sec:consistency:one}

In order to show the consistency of the change-points detected by the SBS algorithm
in terms of their total number and locations, we impose the following assumptions
in addition to (A1).
\begin{itemize}
\item[(A2)] $\{\zktt\}_{t=0}^{T-1}$ is a sequence of standard normal variables and $\max_k\phi^{(k)1}_{\infty}<\infty$, where
\begin{eqnarray*}
\phi^{(k)}(\tau)=\sup_{t,T}|\corr(\zktt, Z^{(k)}_{t+\tau, T})|
\mbox{ \quad and \quad } \phi^{(k)r}_{\infty}=\sum_{\tau}|\phi^{(k)}(\tau)|^r.
\end{eqnarray*}
\item[(A3)] There exist constants $\sig^*, \sig_*>0$ such that $\max_{k,t,T}\sigkt \le \sig^*$,
and given any change-point $\eta_q$ in $\sigkt$,
\begin{eqnarray*}
\left\vert\sig^{(k)}\left(\frac{\eta_q+1}{T}\right)-\sig^{(k)}\left(\frac{\eta_q}{T}\right)\right\vert > \sig_*
\end{eqnarray*}
uniformly for all $k=1, \ldots, d$.
\item[(A4)] $d$ and $T$ satisfy $d \cdot T^{-\log\,T} \to 0$.
\end{itemize}

In particular, condition (A4) specifies the maximum rate at which the dimensionality
$d$ of model (\ref{generic:one}) is permitted to increase with the sample size $T$.
Denoting the estimated change-points (sorted in increasing order) by $\heta_q, \ q=1, \ldots, \wh{N}$,
we have the following result.

\begin{thm}
\label{thm:one}
Let $\Delta_T \asymp \ept$ in the SBS algorithm.
Under (A1)--(A4), there exists $C_1>0$ such that $\heta_q, \ q=1, \ldots, \wh{N}$ satisfy
\begin{eqnarray}
\p\left\{\wh{N}=N; \,|\heta_q-\eta_q| < C_1\ept \mbox{ for } q=1, \ldots, N\right\} \to 1
\nonumber
\end{eqnarray}
as $T \to \infty$, where
\begin{itemize}
\item if $\delt \asymp T$, there exists some positive constant $\kappa$ such that
we have $\ept=\log^{2+\vartheta}\,T$ with $\thr = \kappa\log^{1+\omega}\,T$ for any positive constants $\vartheta$ and $\omega > \vartheta/2$.
\item if $\delt \asymp T^\Theta$ for $\Theta \in (3/4, 1)$,
we have $\ept=T^\theta$ for $\theta=2-2\Theta$ with $\thr=\kappa T^\gamma$ for some $\kappa > 0$ and any $\gamma\in(1-\Theta, \Theta-1/2)$.
\end{itemize}
\end{thm}
We may define the optimality in change-point detection as
when each of the true change-points and the corresponding estimated change-point are within the distance of $O_p(1)$, see e.g. \citet{korostelev1987}.
In this sense, when $\delt \asymp T$, the rate of $\ept$ is near-optimal up to a logarithmic factor.

\subsection{Post-processing of the change-points}
\label{sec:within}

We further equip the SBS algorithm with an extra step aimed at reducing the risk of over-estimating the number of change-points. The step is completely analogous to the corresponding step in the UBS algorithm (see \citet{cho2012},
Section 3.2.1), except it now involves checks of the form
\begin{eqnarray}
\exists\,k\quad\cY^{(k)}_{\heta_{q-1}+1, \heta_q, \heta_{q+1}} > \thr,
\label{eq:within}
\end{eqnarray}
with the convention $\heta_0=0$, $\heta_{\wh{N}+1}=T-1$.
In other words, we compute the CUSUM statistic $\cY^{(k)}_{\cdot, \cdot, \cdot}$ on each triple
of neighboring change-point estimates for each $k$ and only retain those $\heta_q$'s for
which that statistic exceeds the threshold $\thr$ for at least one $k$.
The reader is referred to the above work for
details. As in the UBS algorithm, the consistency result of Theorem \ref{thm:one} is preserved even after performing this extra post-processing.

\section{The SBS algorithm in the multivariate LSW model}
\label{sec:link}

In this section, we demonstrate how the SBS algorithm can be used
for detecting multiple change-points in the second-order (i.e. auto-covariance and cross-covariance)
structure of multivariate, possibly high-dimensional time series.

For this purpose, we first define the multivariate LSW model,
in which wavelets act as building blocks analogous to the Fourier exponentials
in the classical Cram\'{e}r representation for stationary processes.
Our choice of the LSW model as the theoretical setting
is motivated by the attractive features of the univariate LSW model,
listed in \cite{cho2012}.

As the simplest example of a wavelet system, we consider Haar wavelets defined as
\begin{eqnarray}
\psi^H_{i, k} = 2^{i/2}\bbI(0 \le k \le 2^{-i-1}-1) - 2^{i/2}\bbI(2^{-i-1} \le k \le 2^{-i}-1),
\nonumber
\end{eqnarray}
where $i\in\{-1, -2, \ldots\}$ and $k \in \mathbb{Z}$ denote scale and location parameters, respectively.
Small negative values of the scale parameter $i$ denote ``fine'' scales
where the wavelet vectors are the most localized and oscillatory, while
large negative values denote ``coarser'' scales with longer, less oscillatory wavelet vectors.
For a more detailed introduction to wavelets, see e.g. \citet{nason1995} and \citet{vidakovic1999}.
With such wavelets as building blocks, we define the $p$-variate, piecewise stationary LSW model as follows.

\vspace{10pt}

\begin{defn}
\label{def:lsw:one}
The $p$-variate LSW process $\{\bxtt = (X^{(1)}_{t, T}, \ldots, X^{(p)}_{t, T})'\}_{t=0}^{T-1}$
for $T=1, 2, \ldots$, is a triangular stochastic array with the following representation:
\begin{eqnarray}
\xjtt = \sumi\sumk W^{(j)}_i(k/T)\psi_{i, t-k}\xi^{(j)}_{i, k}  \mbox{ \quad for each \ } j=1, \ldots, p,
\label{eq:lsw}
\end{eqnarray}
where $\bxi_{i, k} = (\xi_{i, k}^{(1)}, \xi_{i, k}^{(2)}, \ldots, \xi_{i, k}^{(p)})'$ are
independently generated from multivariate normal distributions
$\cN_p\left(\bzero, \bSig_i(k/T)\right)$, with $\Sigma^{(j, j)}_i(k/T) \equiv 1$ and
\begin{eqnarray*}
\cov(\xi^{(j)}_{i, k}, \xi^{(l)}_{i', k'}) = \left\{
\begin{array}{ll}
\delta_{i, i'}\delta_{k, k'}\cdot\Sigma^{(j, j)}_i(k/T)=\delta_{i, i'}\delta_{k, k'} & \mbox{when } j=l,
\nonumber \\
\delta_{i, i'}\delta_{k, k'}\cdot\Sigma^{(j, l)}_i(k/T) & \mbox{when } j \ne l.
\nonumber
\end{array}\right.
\end{eqnarray*}
The parameters $i\in\{-1, -2, \ldots\}$ and $k \in \mathbb{Z}$ denote scale and location, respectively,
and the Kronecker delta function $\delta_{i, i'}$ returns 1 when $i=i'$ and 0 otherwise.
For each $i$ and $j, l=1, \ldots, p$, the functions $W^{(j)}_i(k/T):[0, 1] \to \mathbb{R}$ and
$\Sigma^{(j, l)}_i(k/T):[0, 1] \to \mathbb{R}$ are piecewise constant
with an unknown number of change-points, and we denote the sets of change-points as
\begin{eqnarray*}
\bbB^{(j)}_i &=& \{z\in(0, 1): \ \lim_{u \to z-}W^{(j)}_i(u) \ne \lim_{u \to z+}W^{(j)}_i(u)\}, \mbox{\quad and} \\
\bbB^{(j, l)}_i &=& \{z\in(0, 1): \ \lim_{u \to z-}\Sigma^{(j, l)}_i(u) \ne \lim_{u \to z+}\Sigma^{(j, l)}_i(u)\}.
\end{eqnarray*}
\end{defn}
\vspace{2mm}
In comparison to the Cram\'{e}r representation for stationary processes,
the functions $W^{(j)}_i(k/T)$ can be thought of as scale- and location-dependent transfer functions,
while the wavelet vectors $\boldsymbol{\psi_i}$ can be thought of
as building blocks analogous to the Fourier exponentials.

The autocovariance and the cross-covariance functions of $\xjtt, \ j=1, \ldots, p$,
defined in Section \ref{sec:sub:wavper} below, inherit the piecewise-constancy of
$W^{(j)}_i(\cdot)$ and $\Sigma^{(j, l)}_i(\cdot)$, with identical change-point
locations. We denote the set of those change-points by
\begin{eqnarray}
\bbB = \left\{\cup_{j=1}^p \bbB^{(j)}\right\} \cup \left\{\cup_{j, l=1}^p \bbB^{(j, l)}\right\} \equiv
\{\nu_r, \ r=1, \ldots, N\}.
\label{lsw:br:set}
\end{eqnarray}

\subsection{Wavelet periodograms and cross-periodograms}
\label{sec:wavper}

In this section, we construct particular wavelet-based local periodogram sequences
from the LSW time series $\bxtt$ in (\ref{eq:lsw}), to which the SBS algorithm
of Section \ref{sec:malg} will be
applied in order to detect the change-points in the second-order structure of $\bxtt$.

Recall that in examples (A)--(B) of Section \ref{sec:malg}, the multiplicative sequences were constructed
as $\yktt = 2^{-1}(X^{(k)}_{t+1, T} - \xtt^{(k)})^2$. Note that each element of $\yktt$ is
simply the squared wavelet coefficient of $\xtt^{(k)}$ with respect to Haar wavelets at
scale $-1$, i.e.
\begin{eqnarray}
\yktt = 2^{-1}(\xtt^{(k)} - X^{(k)}_{t-1, T})^2 = \left(\sum_u X^{(k)}_{u, T}\psi^H_{-1, t-u}\right)^2,
\nonumber
\end{eqnarray}
or the (Haar) \emph{wavelet periodogram} of $\xtt^{(k)}$ at scale $-1$.
In the two examples, it was shown that the change-points in the AR coefficients
of $\xtt^{(k)}$ (and hence in its second-order structure) were detectable
from the wavelet periodograms. In this section, we study the properties of the
wavelet periodogram and cross-periodogram sequences, and discuss the applicability
of the SBS algorithm to the segmentation of $\bxtt$ defined as (\ref{eq:lsw}),
with the wavelet periodograms and cross-periodograms of $\bxtt$ as an input.

\subsubsection{Definitions and properties}
\label{sec:sub:wavper}

Given a $p$-variate LSW time series $\bX_{t, T}=(\xtt^{(1)}, \ldots, \xtt^{(p)})'$,
its empirical wavelet coefficients at scale $i$ are denoted by
$w^{(j)}_{i, t, T}=\sum_u X^{(j)}_{u, T}\psi_{i, t-u}$ for each $\xjtt, \ j=1, \ldots, p$.
Then, the {\em wavelet periodogram}
of $\xjtt$ and the {\em wavelet cross-periodogram} between $\xjtt$ and $\xtt^{(l)}$ at scale $i$
are defined as
\begin{eqnarray*}
I^{(j, j)}_{i, t, T} \equiv \ijitt = \vert w^{(j)}_{i, t, T} \vert^2 \mbox{\quad and \quad}
I^{(j, l)}_{i, t, T} = w^{(j)}_{i, t, T} \cdot w^{(l)}_{i, t, T},
\end{eqnarray*}
respectively. The Gaussianity of $\xjtt$ implies the Gaussianity of $w^{(j)}_{i, t, T}$, and hence
$\ijitt$ and $I^{(j, l)}_{i, t, T}$ admit the following decompositions:
\begin{eqnarray}
\ijitt &=& \E \ijitt \cdot \zjstt, \quad t=0, \ldots, T-1, \label{waveper} \\
I^{(j, l)}_{i, t, T} &=& \E I^{(j, l)}_{i, t, T} \cdot Z^{(j)}_{t, T}Z^{(l)}_{t, T}, \quad t=0, \ldots, T-1, \label{wavecrossper}
\end{eqnarray}
where $\{\ztt^{(j)}\}_{t=0}^{T-1}$ is a sequence of (correlated and nonstationary) standard normal variables for each $j=1, \ldots, p$.
Therefore each $\ijitt$ follows a scaled $\chi^2_1$ distribution.

It has been shown in the literature that for a univariate LSW process $\xtt$,
there exists an asymptotic one-to-one correspondence
between its time-varying autocovariance functions
$c_T(z, \tau) = \cov(X_{\lfloor zT\rfloor, T}, X_{\lfloor zT\rfloor+\tau, T}), \ \tau=0, 1, \ldots$,
transfer functions $W_i^2(z)$, and the expectations of wavelet periodograms $\E I_{i, t, T}$ at multiple scales (see e.g. \citet{cho2012}).
That is, any change-points in the set of piecewise constant functions $\{W_i^2(z)\}_i$ correspond to
change-points in the (asymptotic limits of the) autocovariance functions $\{c_T(z, \tau)\}_{\tau}$,
which in turn correspond to the change-points in the (asymptotic limits of the) functions
$\{\E I_{i, t, T}\}_i$, and thus are asymptotically detectable by examining $I_{i, t, T}, \ i=-1, -2, \ldots$.
For a multivariate LSW process $\bxtt$, its autocovariance and cross-covariance functions are defined as
\begin{eqnarray}
c^{(j, j)}_T(z, \tau) = c^{(j)}_T(z, \tau) =
\cov(X^{(j)}_{\lfloor zT\rfloor, T}, X^{(j)}_{\lfloor zT\rfloor+\tau, T}) \mbox{ \ and \ }
c^{(j, l)}_T(z, \tau) = \cov(X^{(j)}_{\lfloor zT\rfloor, T}, X^{(l)}_{\lfloor zT\rfloor+\tau, T}).
\end{eqnarray}
In the multivariate LSW model, analogous one-to-one correspondence can be shown
for any pair of $\xjtt$ and $X^{(l)}_{t, T}$ between the following quantities:
the autocovariance and cross-covariance functions $c^{(j)}_T(z, \tau)$,
$c^{(l)}_T(z, \tau)$ and $c^{(j, l)}_T(z, \tau)$ at lags $\tau = 0, 1, \ldots$,
piecewise constant functions $\{W^{(j)}_i(z)\}^2$, $\{W^{(l)}_i(z)\}^2$ and $\Sigma^{(j, l)}_i(z)$,
and the expectations of wavelet periodograms and cross-periodograms $\E\ijitt$, $\E I^{(l)}_{i, t, T}$
and $\E I^{(j, l)}_{i, t, T}$ at scales $i=-1, -2, \ldots$.
Therefore, any change-points in the second-order structure of the multivariate time series $\bxtt$
are detectable from the wavelet periodograms and cross-periodograms at multiple scales.
Formal derivation of this one-to-one correspondence is provided in Appendix \ref{app:two}.

Thus we now focus on wavelet periodogram $\ijitt$ and cross-periodogram $I^{(j, l)}_{i, t, T}$ as the input to the SBS algorithm.
We firstly note that $\E\ijitt$ are piecewise constant except for negligible biases around the change-points
(which are accounted for in our results, see Section B.1 in the Appendix),
and thus $\ijitt$ ``almost'' follow the multiplicative model (\ref{generic:one}).
However, $I^{(j, l)}_{i, t, T}$ is not of the form specified in
(\ref{generic:one}) and the next section introduces an alternative
to $I^{(j, l)}_{i, t, T}$ which does follow (\ref{generic:one}) (again,
up to the negligible biases) and contains the same information
about the change-points as does $I^{(j, l)}_{i, t, T}$.

\subsubsection{Non-negative multiplicative alternative to the cross-periodogram}
\label{sec:alt:wavecrossper}

\begin{figure}[htbp]
\centering
\epsfig{file=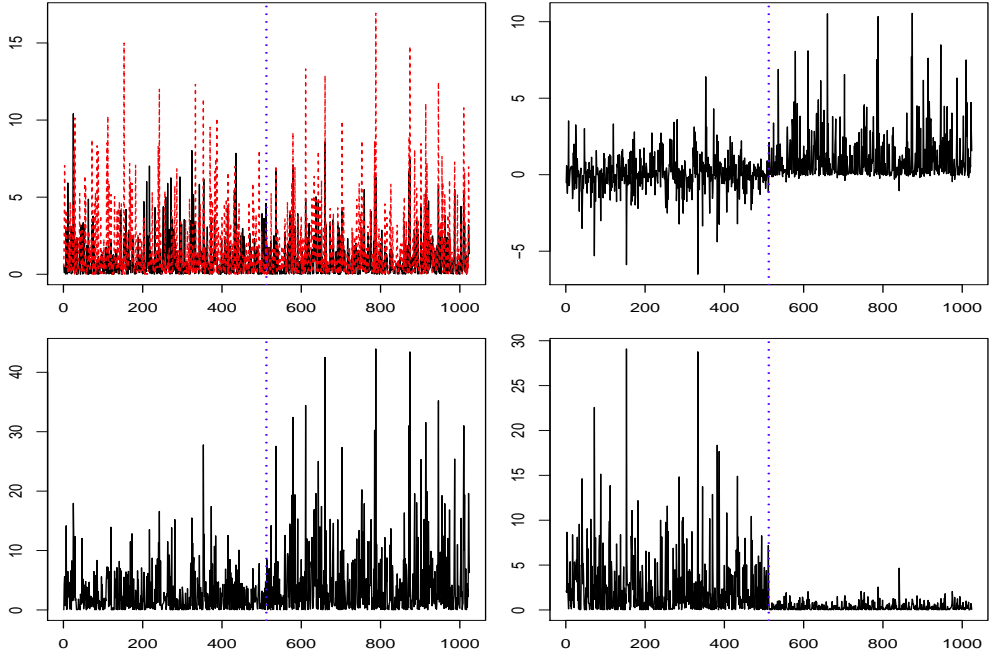, width=.8\linewidth, height=3.5in}
\caption{
Top left: $a_t^2$ (solid) and $b_t^2$ (broken) from the example of Section \ref{sec:alt:wavecrossper};
top right: $a_t \cdot b_t$;
bottom left: $(a_t+b_t)^2$; bottom right $(a_t-b_t)^2$;
dotted vertical lines denote where $(a_t, b_t)'$ have change-points.}
\label{fig:ab}
\end{figure}

To gain an insight into obtaining a possible alternative to $I^{(j, l)}_{i, t, T}$,
we first present a toy example. Consider two sequences of zero-mean, serially independent normal variables
$\{a_t\}_{t=1}^T$ and $\{b_t\}_{t=1}^T$ where
the correlation between $a_t$ and $b_t$ satisfies $\corr(a_t, b_t) = 0$ for $t \le \lfloor T/2 \rfloor$
and $\corr(a_t, b_t)=0.9$ for $t \ge \lfloor T/2 \rfloor +1$, while $\var(a_t)$ and $\var(b_t)$ are constant
over time. The change in the second-order structure of $(a_t, b_t)'$ originates solely
from the change in the correlation between the two sequences,
and thus cannot be detected from $\{a_t^2\}_{t=1}^T$ and $\{b_t^2\}_{t=1}^T$ alone.
Figure \ref{fig:ab} confirms this, and
it is the sequence $\{(a_t-b_t)^2\}_{t=1}^T$ that exhibits the change-point
more prominently than $\{a_tb_t\}_{t=1}^T$ or $\{(a_t+b_t)^2\}_{t=1}^T$.

Identifying $a_t$ with $w^{(j)}_{i, t, T}$ and $b_t$ with $w^{(l)}_{i, t, T}$,
it becomes apparent that we may detect any change in the covariance structure between
$w^{(j)}_{i, t, T}$ and $w^{(l)}_{i, t, T}$
by examining $\ijitt$, $I^{(l)}_{i, t, T}$, and either
$(w^{(j)}_{i, t, T} + w^{(l)}_{i, t, T})^2$ or $(w^{(j)}_{i, t, T} - w^{(l)}_{i, t, T})^2$
instead of $I^{(j, l)}_{i, t, T} = w^{(j)}_{i, t, T} w^{(l)}_{i, t, T}$.
Since each variable $w^{(j)}_{i, t, T}$ is zero-mean normal, both $(w^{(j)}_{i, t, T} \pm w^{(l)}_{i, t, T})^2$
are scaled $\chi^2_1$ variables, and so either of these sequences
can serve as an input to the SBS algorithm. While both lead to identical results theoretically,
there remains the choice between the signs $\pm$ to optimize finite-sample performance.
Our empirical
observation is that the choice
\begin{eqnarray}
\tilde{I}^{(j, l)}_{i, t, T} = \left(w^{(j)}_{i, t, T}-\mbox{sign}\left(\wh{\corr}(w^{(j)}_{i, t, T}, w^{(l)}_{i, t, T})\right)
\cdot w^{(l)}_{i, t, T}\right)^2,
\label{cross}
\end{eqnarray}
where $\wh{\corr}(\cdot, \cdot)$ is the sample correlation computed separately on each
current segment, performs well, and we adopt it in practice.
In summary, the multiplicative sequences that comprise the input to the SBS algorithm are
$\ijitt$ and $\tilde{I}^{(j, l)}_{i, t, T}$ for $j, l=1, \ldots, p$.

\subsection{Application of the SBS algorithm to multivariate time series}
\label{sec:consistency:two}

We expect $\ijitt$ ($\tilde{I}^{(j, l)}_{i, t, T}$) at finer scales to provide more accurate information
on the presence and locations of the change-points in $\E\ijitt$ ($\E\tilde{I}^{(j, l)}_{i, t, T}$),
while those at coarser scales to be of limited use. This is due to the increasing length
$\mathcal{L}_i$ of the support of the wavelet vectors $\bps_i$ at coarser scales, as well as
the resulting increasing autocorrelation in $\{w^{(j)}_{i, t, T}\}_{t=0}^{T-1}$.
In addition, since the number of periodogram and cross-periodogram sequences increases
by $p(p+1)/2$ with each scale added, limiting the number of scales also carries clear computational
benefits, especially in high dimensions.
Therefore we propose to consider $\ijitt$ and $\tilde{I}^{(j, l)}_{i, t, T}$ scale by scale,
starting from the finest scale $i=-1$ and
ending with scale $\is_T = -\lfloor\alpha\log\log\,T\rfloor$
with $\alpha\in(0, 2+\vartheta]$, with the latter choice being made
to guarantee consistency of our procedure.

Having detected the change-points at each scale separately,
we then reduce the set of estimated change-points such that
those estimated on different scales yet indicating the same change-point, are combined into one
with high probability. This is done in the same way as in the univariate case
and is described in detail in \citet{cho2012}. Here, we only mention that
this across-scales post-processing procedure involves a parameter $\Lambda_T$ which determines the
maximum diameter of the initial clusters of change-points originating from
different scales.

Summarizing the above arguments, we propose the following algorithm for the
segmentation of multivariate time series with piecewise constant second-order structure.
We label it SBS-MVTS (Sparsified Binary Segmentation for MultiVariate
Time Series). Its core ingredient is the SBS algorithm, described in Section \ref{sec:malg}.

\begin{description}
\item[SBS-MVTS algorithm]
\item[Step 0] Set the scale parameter to $i = -1$ (the finest scale).
\item[Step 1] Apply the SBS algorithm as well as the post-processing step
of Section \ref{sec:within} to the $d \equiv p(p+1)/2$ sequences
$\ijitt, \ j=1, \ldots, p$ and $\tilde{I}^{(j, l)}_{i, t, T}, \ j \ne l; \,j, l=1, \ldots, p$,
and denote the detected change-points by $\hnu_{i, r}, \ r=1, \ldots, \wh{N}_i$.
\item[Step 2] Update $i \leftarrow i-1$ and repeat Step 1 until $i$ reaches $\is_T$.
Apply the across-scales post-processing (described earlier in this section) to
the change-points $\hnu_{i, r}, \ r=1, \ldots, \wh{N}_i$ detected from the scales $i=-1, \ldots, \is_T$,
and obtain the final set of estimated change-points $\hnu_r, \ r=1, \ldots, \wh{N}$.
\end{description}
The following theorem demonstrates that the consistency of the SBS algorithm
for the multiplicative sequences in (\ref{generic:one}) carries over to that
of the SBS-MVTS algorithm, provided that the $p$-variate LSW time series $\bxtt$ on input
satisfies conditions (B1)--(B5) (in Appendix \ref{app:two}), which are analogues of conditions (A1)--(A4)
but phrased in the specific context of LSW processes. In particular, condition
(B5) states that the dimensionality $p$ of the input time series $\bxtt$ is permitted to increase
with $T$ as long as $p^2 T^{-\log\,T} \to 0$.

\vspace{10pt}

\begin{thm}
\label{thm:two}
Let $\Delta_T \asymp \ept$ in the SBS algorithm and $\Lambda_T \asymp \ept$ in the across-scales post-processing.
Under (B1)--(B5), there exists $C_2>0$ such that $\hnu_r, \ r=1, \ldots, \wh{N}$
estimated with $\is_T = -\lfloor\alpha\log\log\,T\rfloor$ for $\alpha\in(0, 2+\vartheta]$, satisfy
\begin{eqnarray}
\p\left\{\wh{N}=N; \ |\hnu_r-\nu_r| < C_2\ept \mbox{ for } r=1, \ldots, N\right\} \to 1
\nonumber
\end{eqnarray}
as $T \to \infty$, where
\begin{itemize}
\item if $\delt \asymp T$, there exists some positive constant $\kappa$ such that
we have $\ept=\log^{2+\vartheta}\,T$ with $\thr = \kappa\log^{1+\omega}\,T$ for any positive constants $\vartheta$ and $\omega > \vartheta/2$.
\item if $\delt \asymp T^\Theta$ for $\Theta \in (3/4, 1)$,
we have $\ept=T^\theta$ for $\theta=2-2\Theta$ with $\thr=\kappa T^\gamma$ for some $\kappa > 0$ and any $\gamma\in(1-\Theta, \Theta-1/2)$.
\end{itemize}
\end{thm}

\subsection{Practical choice of threshold and other quantities}
\label{sec:thr}

The aim of this section is to provide some practical guidance as
to the choice of various parameters of the SBS-MVTS algorithm.
We provide heuristic justification for the chosen values below.
They have been found to work well in our extensive simulation
studies across a range of models; however, we do not claim that
other values would not work equally well or better in practice.

Importantly, we also note that the necessity of calibrating these parameters is not
specific to the SBS-MVTS algorithm in the sense that they would
also need to be set if, for example, $\ty^{\mbox{\scriptsize avg}}_t$ or
$\ty^{\mbox{\scriptsize max}}_t$ were used instead of
$\ty^{\mbox{\scriptsize thr}}_t$ in a binary segmentation framework.

From the conditions of Theorem \ref{thm:one}, we have $\gamma\in(1-\Theta, \Theta-1/2)$
in the threshold $\thr = \kappa T^\gamma$ when $\Theta\in(3/4, 1)$,
while $\omega$ is any positive constant greater than $\vartheta/2$
in $\thr = \kappa\log^{1+\omega}\,T$ when $\Theta=1$.
We propose to set $\gamma$ as conservatively as $\gamma=0.499$
and focus on the choice the constant $\kappa$ for each $\xjtt$,
by simulating wavelet periodograms under the null hypothesis of no change-points as below.
With this approach to the selection of $\kappa$, 
finite sample performance is little affected by
whether $T^\gamma$ or $\log^{1+\omega}\,T$ is used as the rate of $\thr$,
and thus we do not expand on the choice of $\omega$ here.

For each univariate process $\xjtt$, we estimate $\wh{a}_j$, its lag-one autocorrelation.
Then, generating AR(1) time series of length $T$ with the AR parameter $\wh{a}_j$
repeatedly $R$ times, we compute the following statistic for each realization $m$:
\begin{eqnarray}
\bbJ^{(j, m)}_i=\max_t \left(\frac{1}{T}\sum_{u=1}^T I^{(j, m)}_{i, u}\right)^{-1}
\left\vert\sqrt{\frac{T-t}{T\cdot t}}\sum_{u=1}^t I^{(j, m)}_{i, u}
-\sqrt{\frac{t}{T(T-t)}}\sum_{u=t+1}^T I^{(j, m)}_{i, u} \right\vert,
\nonumber
\end{eqnarray}
where $I^{(j, m)}_{i, t}$ denotes the scale $i$ wavelet periodogram of the $m$th AR(1) process
generated with the AR parameter $\wh{a}_j$.
Note that $\bbJ^{(j, m)}_i$ is of the same form as the test statistic used in the SBS algorithm.
Since the AR processes have been generated under the null hypothesis of no change-points in their second-order structure,
$T^{-\gamma} \bbJ^{(j, m)}_i$ may serve as a proxy for $\kappa$ for the wavelet periodograms generated from $\xjtt$.
We have observed that the values of $\bbJ^{(j, m)}_i$ tend to increase at coarser scales
due to the increasing support of the wavelet vector $\boldsymbol{\psi}_i$.
Therefore, we select $\kappa$ to be scale-dependent as $\kappa^{(j)}_i$
for each $i=-1, -2, \ldots$ and $j=1, \ldots, p$. In the SBS algorithm,
we choose it to be the $99\%$-quantile of $T^{-\gamma} \bbJ^{(j, m)}_i$
over all $m=1, \ldots, R$. In the case of wavelet cross-periodograms, we use the
first-lag sample autocorrelation of $\xjtt-\mbox{sign}\{\wh{\corr}(w^{(j)}_{i, t, T},
w^{(l)}_{i, t, T})\}X^{(l)}_{t, T}$ in place of $\wh{a}_j$.

As for the choice of $\Delta_T$ in Step 2.3 of the SBS algorithm,
since $\Delta_T \asymp \ept$, we choose $\Delta_T = \lfloor \sqrt{T}/2 \rfloor$ to be on conservative side
and use it in our implementation for the simulations study reported in the next section.
Also we use $\alpha = 2$ and $\Lambda_T = \lfloor \sqrt{T}/2 \rfloor$.
Finally, rather than choosing a fixed constant as $c_*$,
we make sure that a newly detected change-point is distanced from
the previously detected change-points by at least $\Delta_T$.

\section{Simulation study}
\label{sec:sim}

In this section, we study the performance of the SBS-MVTS algorithm on
simulated multivariate time series
with time-varying second-order structure.
All simulated datasets are generated with $T=1024$, and
the sparsity of the change-points across the $p$-dimensional time series is controlled
such that $\lfloor \varrho p\rfloor$ processes out of the $p$ have at least one change-point,
from a sparse case ($\varrho=0.05$) through moderate cases ($\varrho=0.25, 0.5$) to a dense case ($\varrho=1$).
\begin{description}
\item[(M1) Autoregressive (AR) time series.] \hfill \\
We simulate the $p$ time series as AR(1) processes
\begin{eqnarray}
X^{(j)}_t = \alpha^{(j)}X^{(j)}_{t-1}+\sigma^{(j)}\ep_t^{(j)}, \quad j=1, \ldots, p.
\label{sim:one}
\end{eqnarray}
The AR coefficients are independently generated from the uniform distribution
$\cU(-0.5, 0.999)$, and $\sigma^{(j)}$ from $\cU(1/2, 2)$.
The error terms $\bep_t=(\ep^{(1)}_t, \ldots, \ep^{(p)}_t)'$ are generated from $\cN_p(\bzero, \bSig_\ep)$
with $\bSig_\ep$ specified below.
There are three change-points located at $t=341, 614, 838$ which occur in the following ways.
\begin{description}
\item[(M1.1)]
At each change-point, both $\alpha^{(j)}$ and $\sigma^{(j)}$ are re-generated
for randomly chosen $\lfloor \varrho p\rfloor$ time series $X^{(j)}_t$, while
$\bSig_\ep = 4\cdot\bI_p$ and remains unchanged throughout.

\item[(M1.2)]
Originally, $\bep_t$ is generated with a block-diagonal variance-covariance
matrix $\bSig_\ep = (\Sigma_{j, l})_{j, l=1}^p$, where
$\Sigma_{j,j} = 4$ for $j = 1, \ldots, p$, and $\Sigma_{j, l}=4(-0.95)^{|j-l|}$ for
$j, l=1, \ldots, p/2$ and zero elsewhere.
The cross-correlation structure of $\bep_t$ changes at each change-point
as the locations of randomly chosen $\lfloor \varrho p/2\rfloor$ elements of $\bep_t$ are swapped
with those of other $\lfloor \varrho p/2\rfloor$ randomly chosen elements on each stationary segment.
\end{description}

This model has been chosen for the simplicity of the AR(1) dependence structure and
for the fact that it permits easy manipulation of the cross-dependence between the
component series.

\item[(M2) Factor models.] \hfill \\
The $p$ time series are generated from a factor model
\begin{eqnarray}
\bX_t = \bA\bet_t+\bvep_t,
\nonumber
\end{eqnarray}
where $\bA$ is a $p \times 5$ factor loading matrix with each element $A_{j, l}$ generated from a uniform distribution
$\cU(0.5, 1.5)$. The vector $\bet_t$ contains five factors, each of which is an independent AR(1) time series
generated as $X^{(j)}_t$ in (\ref{sim:one}) with $\bSig_\ep = 4\cdot\bI_p$.
The error terms $\bvep_t$ follow $\cN_p(\bzero, \bSig_\vep)$ with the same covariance matrix as that in (M1.2).
There are three change-points located at $t=341, 614, 838$ which occur in the following ways.
\begin{description}
\item[(M2.1)] At each change-point, $\lfloor \varrho p\rfloor$ randomly chosen rows of the factor loading matrix $\bA$
are re-generated, each from $\cN(0, 1)$.
\item[(M2.2)] The cross-correlation structure of $\bvep_t$ changes as in (M1.2).
\end{description}

The aim of this model is to investigate the performance of our algorithm
when the dependence structure is governed by a factor model, a popular
dimensionality reduction tool for high-dimensional time series.

\item[(M3) AR(1)$+$MA(2) model.] \hfill \\
In this example, the $p$-variate time series $\bX_t$ is generated such that,
\begin{eqnarray}
X^{(j)}_t = \left\{\begin{array}{ll}
\ep_t^{(j)}+\beta_1^{(j)}\ep_{t-1}^{(j)}+\beta_2^{(j)}\ep_{t-2}^{(j)}  & \mbox{for } 1 \le t \le 512, \\
\alpha^{(j)}X_t^{(j)}+\sigma^{(j)}\ep_t^{(j)} & \mbox{for } 513 \le t \le 1024,
\end{array}\right.
\nonumber
\end{eqnarray}
for $j=1, \ldots, \lfloor\varrho p\rfloor$, and $X^{(j)}_t, \ j = \lfloor\varrho p\rfloor+1, \ldots, p$
are stationary AR(1) processes with the AR parameters generated from $\cU(-0.5, 0.999)$
and $\var(\ep_{t}^{(j)}) = 1$.
The coefficients $\beta_1^{(j)}$, $\beta_2^{(j)}$, $\alpha^{(j)}$ and $\sigma^{(j)}$ are generated such that
for $X^{(j)}_t, \ j=1, \ldots, \lfloor\varrho p\rfloor$,
the variance and the first-lag autocorrelation remain constant before and after the change-point at $t=512$,
while autocorrelations at other lags have a change-point at $t=512$.
The purpose of this model is to investigate whether the SBS-MVTS algorithm can perform well
when the change-points are not detectable at the finest scale $i=-1$.

\item[(M4) Short segment.] \hfill \\
Inspired by the example (B) of Section \ref{sec:malg}, the $p$-variate time series $\bX_t$ is generated such that
the first $\lfloor\varrho p\rfloor$ processes follow
\begin{eqnarray}
X^{(j)}_t = \left\{\begin{array}{ll}
\alpha^{(j)}X_t^{(j)}+\ep_t^{(j)} & \mbox{for } 1 \le t \le 100, \\
\beta^{(j)}X_t^{(j)}+\ep_t^{(j)} & \mbox{for } 101 \le t \le 1024,
\end{array}\right.
\nonumber
\end{eqnarray}
with $\alpha^{(j)}$ drawn from $\cU(0.5, 0.59)$ and $\beta^{(j)}$ from $\cU(-0.79, -0.5)$.
The remaining $(p-\lfloor\varrho p\rfloor)$ time series are generated as stationary AR(1) processes
with the AR parameters drawn from the same distribution as $\beta^{(j)}$.
The purpose of this model is to
investigate if the SBS-MVTS algorithm performs well
when the finest scale wavelet periodograms suffer from high autocorrelation
while at the same time, the two stationary segments defined by the change-point are of substantially different lengths.
\end{description}
Most methods for multivariate time series segmentation proposed in the literature,
such as those cited in the Introduction, have not been designed for data of the
dimensionality or size considered in this paper, which are $p=50, 100$ and $T=1024$, respectively
(recall that $d$ is quadratic in $p$).

In what follows, we compare the performance of the SBS-MVTS algorithm to that of identical
binary segmentation algorithms but constructed using
$\ty^{\mbox{\scriptsize avg}}_t$ and $\ty^{\mbox{\scriptsize max}}_t$
in (\ref{eq:groen}) instead of $\ty^{\mbox{\scriptsize thr}}_t$.
For clarity, in the remainder of this section, we refer to the three
algorithms as THR (=SBS-MVTS), AVG and MAX.
Identical thresholds $\thr$ are applied in the THR and MAX.
As for the AVG, we test $\ty^{\mbox{\scriptsize avg}}_t$ using a scaled threshold
$d^{-1}\sum_{k=1}^d \bbI(\max_{t\in(s, e)}\cY^{(k)}_{s, t, e}>\thr) \cdot \thr$ to ensure fairer comparison.
As an aside, we note that the threshold selection via simulation is easier for the THR and MAX
algorithms than for the AVG algorithm, the reason being that in the former two cases it can
be reduced to the problem of threshold selection for univariate time series, which is not the case
for AVG.

Tables \ref{table:sim:one}--\ref{table:sim:four} report the results of applying the three segmentation algorithms
to the simulated datasets from (M1)--(M4).
Each table reports the mean and standard deviation of the total number of detected change-points
over 100 simulated time series, and the percentage of ``correctly'' identifying each change-point in the time series
(in the sense that it lies within the distance of $\lfloor \sqrt{T}/2 \rfloor$ from the true change-points).

Overall, it is evident that the THR algorithm outperforms the other two.
In particular, the performance of AVG does not match that of THR
or MAX especially when the change-points are sparse: in some of the models,
there is a tendency for AVG to overestimate the number of change-points.
Besides, the standard deviation of the number of change-points detected by
AVG tends to be larger than those for the other two algorithms.

In terms of the {\em number} of detected change-points, THR and MAX perform
similarly well. However, the accuracy of the detected change-point {\em locations}
is significantly better for THR than for MAX, especially in models (M3)--(M4). This is
unsurprising as effectively, the MAX algorithm locates change-points based on
one individual component of the input time series, while THR typically averages
information across many components.
We also note that the performance of the THR algorithm does not differ greatly between the cases
when $p=50$ and when $p=100$.

As noted earlier, the input sequences to the segmentation algorithms,
$\ijitt$ and $\tilde{I}^{(j, l)}_{i, t, T}$, have expectations which are almost piecewise constant
but not completely so, due to negligible biases around the change-points (see Appendix \ref{app:two:two}).
In deriving Theorem \ref{thm:two}, these biases have fully been taken into account,
which implies that the consistency of SBS-MVTS is extended to the case where
changes occur in the second-order structure of $\bX_{t, T}$
within a short period of time (to be precise, of length $C(\log\,T)^\alpha$ for some $C>0$ and $\alpha$ from $\is_T$), but not entirely synchronized.
To confirm this, we performed a further simulation study where the $p$-variate time series was generated from (M3),
except that the change-points were allowed to be anywhere within an interval of length $\lfloor 2\log\,T \rfloor$ around $t=512$.
Although not reported here, we obtained the change-point detection results with $T=1024$ and varying $\varrho$ and $p$,
which were comparable to those reported in Table \ref{table:sim:three}.
More specifically, while the number of detected change-points had greater variance,
the accuracy in their locations was preserved even when the change-points were not aligned.
Also, overall, the THR algorithm still outperformed the two other competitors in terms of both the total number of the detected change-points
and their locations.

(We now abandon the THR notation and revert to the SBS-MVTS notation in the
remainder of the paper.)

\begin{table}
\caption{Summary of the change-points detected from (M1): mean and standard deviation of the total number of detected change-points, and the percentage of correctly identifying each change-point at $t=341, 614, 838$ over 100 simulated time series.}
\label{table:sim:one}
\centering
\scriptsize{
\begin{tabular}{ c | c | c c c | c c c || c c c | c c c}
\hline
\hline
& & \multicolumn{6}{c}{$p=50$}
& \multicolumn{6}{c}{$p=100$} \\
& & \multicolumn{3}{c}{(M1.1)}
& \multicolumn{3}{c}{(M1.2)}
& \multicolumn{3}{c}{(M1.1)}
& \multicolumn{3}{c}{(M1.2)} \\
$\varrho$ &  & THR & AVG & MAX &
THR & AVG & MAX &
THR & AVG & MAX &
THR & AVG & MAX \\
\hline
\multirow{5}{*}{0.05} & mean & 3.03 &	2.61 &	3.01 &	2.81 &	3.78 &	2.8 &	3.06 &	3 &	3.02 &	3.33 &	4.97 &	3.34 \\
 & sd & 0.17 &	0.71 &	0.1 &	0.44 &	1.34 &	0.45 &	0.24 &	0.83 &	0.14 &	0.55 &	1.23 &	0.54 \\
 & $t=341$ & 98 &	71 &	95 &	91 &	65 &	88 &	97 &	55 &	96 &	97 &	55 &	96 \\
 & $t=614$ & 89 &	75 &	92 &	91 &	67 &	92 &	99 &	55 &	91 &	99 &	55 &	91 \\
 & $t=838$ & 92 &	76 &	91 &	93 &	60 &	91 &	94 &	50 &	87 &	94 &	50 &	87 \\
\hline
\multirow{5}{*}{0.25} & mean & 3.03 &	3.23 &	3.07 &	3.01 &	4.8 &	3.03 &	3.08 &	3.27 &	3.14 &	3.02 &	4.92 &	3.01 \\
 & sd & 0.17 &	0.58 &	0.26 &	0.1 &	1.13 &	0.17 &	0.27 &	0.57 &	0.4 &	0.14 &	1.24 &	0.1 \\
 & $t=341$ & 100 &	100 &	86 &	100 &	73 &	89 &	98 &	100 &	84 &	100 &	65 &	87 \\
 & $t=614$ & 89 &	100 &	91 &	100 &	57 &	88 &	89 &	99 &	88 &	100 &	66 &	93 \\
 & $t=838$ & 99 &	99 &	95 &	100 &	55 &	92 &	99 &	100 &	92 &	100 &	66 &	88 \\
\hline
\multirow{5}{*}{0.5} & mean & 3.05 &	3.21 &	3.05 &	3.01 &	4.66 &	3 &	3.15 &	3.48 &	3.24 &	3.04 &	4.9 &	3.06 \\
 & sd & 0.22 &	0.52 &	0.22 &	0.1 &	1.02 &	0 &	0.36 &	0.64 &	0.51 &	0.2 &	1.14 &	0.24 \\
 & $t=341$ & 100 &	100 &	85 &	99 &	70 &	90 &	100 &	100 &	80 &	100 &	67 &	88 \\
 & $t=614$ & 91 &	100 &	83 &	100 &	69 &	88 &	100 &	100 &	82 &	100 &	68 &	91 \\
 & $t=838$ & 98 &	100 &	80 &	100 &	58 &	86 &	100 &	100 &	84 &	100 &	65 &	87 \\
\hline
\multirow{5}{*}{1} & mean & 3.07 &	3.25 &	3.13 &	3.01 &	4.76 &	3.04 &	3.11 &	3.59 &	3.24 &	3.04 &	5.03 &	3.09 \\
 & sd & 0.26 &	0.52 &	0.37 &	0.1 &	1.1 &	0.2 &	0.31 &	0.81 &	0.45 &	0.24 &	1.27 &	0.29 \\
 & $t=341$ & 98 &	100 &	72 &	100 &	61 &	88 &	100 &	100 &	65 &	100 &	75 &	84 \\
 & $t=614$ & 99 &	100 &	82 &	100 &	65 &	83 &	100 &	100 &	79 &	100 &	73 &	88 \\
 & $t=838$ & 100 &	100 &	89 &	100 &	63 &	85 &	100 &	100 &	88 &	100 &	67 &	84 \\
\hline
\end{tabular}}
\end{table}

\begin{table}
\caption{Summary of the change-points detected from (M2).}
\label{table:sim:two}
\centering
\scriptsize{
\begin{tabular}{ c | c | c c c | c c c || c c c | c c c}
\hline
\hline
& & \multicolumn{6}{c}{$p=50$}
& \multicolumn{6}{c}{$p=100$} \\
& & \multicolumn{3}{c}{(M2.1)}
& \multicolumn{3}{c}{(M2.2)}
& \multicolumn{3}{c}{(M2.1)}
& \multicolumn{3}{c}{(M2.2)} \\
$\varrho$ &  & THR & AVG & MAX &
THR & AVG & MAX &
THR & AVG & MAX &
THR & AVG & MAX \\
\hline
\multirow{5}{*}{0.05} & mean & 3.04 &	1.86 &	3.11 &	2.81 &	3.07 &	2.79 &	2.98 &	3.01 &	2.95 &	3.27 &	3.94 &	3.29 \\
 & sd & 0.2 &	0.96 &	0.35 &	0.44 &	1.28 &	0.46 &	0.28 &	1 &	0.3 &	0.57 &	1.04 &	0.59 \\
 & $t=341$ & 91 &	44 &	88 &	91 &	61 &	92 &	90 &	67 &	81 &	99 &	50 &	86 \\
 & $t=614$ & 89 &	35 &	90 &	92 &	55 &	87 &	89 &	52 &	77 &	96 &	44 &	84 \\
 & $t=838$ & 95 &	41 &	85 &	89 &	59 &	83 &	86 &	52 &	79 &	96 &	43 &	88 \\
\hline
\multirow{5}{*}{0.25} & mean & 3.02 &	3.07 &	3.05 &	3 &	4.11 &	3.01 &	3.01 &	3.43 &	3.05 &	3.01 &	4.42 &	3.01 \\
 & sd & 0.14 &	0.76 &	0.22 &	0 &	0.96 &	0.1 &	0.1 &	0.78 &	0.22 &	0.1 &	1.16 &	0.1 \\
 & $t=341$ & 95 &	66 &	93 &	98 &	70 &	95 &	93 &	77 &	80 &	100 &	71 &	88 \\
 & $t=614$ & 95 &	66 &	91 &	100 &	74 &	86 &	90 &	69 &	84 &	100 &	79 &	92 \\
 & $t=838$ & 93 &	58 &	85 &	100 &	70 &	93 &	94 &	74 &	84 &	100 &	70 &	91 \\
\hline
\multirow{5}{*}{0.5} & mean & 3.01 &	3.07 &	3.02 &	3 &	4.43 &	3.02 &	3.03 &	3.11 &	3.07 &	3 &	4.32 &	3.04 \\
 & sd & 0.1 &	0.48 &	0.14 &	0 &	1.08 &	0.14 &	0.17 &	0.31 &	0.26 &	0 &	1.07 &	0.2 \\
 & $t=341$ & 97 &	82 &	86 &	97 &	75 &	85 &	96 &	88 &	83 &	100 &	79 &	93 \\
 & $t=614$ & 93 &	80 &	87 &	100 &	76 &	94 &	92 &	98 &	77 &	100 &	73 &	89 \\
 & $t=838$ & 93 &	90 &	80 &	98 &	67 &	87 &	95 &	95 &	82 &	99 &	70 &	93 \\
\hline
\multirow{5}{*}{1} & mean & 3 &	3.1 &	3.01 &	3.01 &	4.05 &	3.02 &	3 &	3.2 &	3.03 &	3 &	4.27 &	3.05 \\
 & sd & 0 &	0.36 &	0.1 &	0.1 &	1.08 &	0.14 &	0 &	0.57 &	0.17 &	0 &	1.12 &	0.22 \\
 & $t=341$ & 99 &	71 &	94 &	99 &	71 &	94 &	94 &	100 &	84 &	98 &	70 &	86 \\
 & $t=614$ & 99 &	72 &	89 &	99 &	72 &	89 &	93 &	100 &	89 &	100 &	78 &	86 \\
 & $t=838$ & 100 &	69 &	91 &	100 &	69 &	91 &	94 &	100 &	81 &	100 &	66 &	80 \\
\hline
\end{tabular}}
\end{table}

\begin{table}
\caption{Summary of the change-points detected from (M3).}
\label{table:sim:three}
\centering
\scriptsize{
\begin{tabular}{ c | c | c c c || c c c }
\hline
\hline
& & \multicolumn{3}{c}{$p=50$}
& \multicolumn{3}{c}{$p=100$} \\
$\varrho$ &  & THR & AVG & MAX &
THR & AVG & MAX \\
\hline
\multirow{3}{*}{0.05} & mean & 0.63 &	0.03 &	0.69 &	1.08 &	0.09 &	0.98 \\
 & sd & 0.53 &	0.17 &	0.49 &	0.46 &	0.32 &	0.62 \\
 & $t=512$ & 51 &	1 &	49 &	72 &	3 &	64 \\
\hline
\multirow{3}{*}{0.25} & mean & 1.01 &	0.11 &	1.04 &	1.02 &	0.32 &	0.99 \\
 & sd & 0.22 &	0.31 &	0.28 &	0.2 &	0.51 &	0.33 \\
 & $t=512$ & 92 &	9 &	73 &	92 &	25 &	75 \\
\hline
\multirow{3}{*}{0.5} & mean & 1.01 &	0.31 &	1 &	1.05 &	0.47 &	1.07 \\
 & sd & 0.17 &	0.51 &	0.28 &	0.22 &	0.56 &	0.29 \\
 & $t=512$ & 92 &	29 &	77 &	90 &	39 &	68 \\
\hline
\multirow{3}{*}{1} & mean & 1.02 &	0.36 &	1.03 &	1.13 &	0.68 &	1.22 \\
 & sd & 0.14 &	0.5 &	0.22 &	0.34 &	0.65 &	0.46 \\
 & $t=512$ & 95 &	34 &	67 &	95 &	53 &	66 \\
\hline
\end{tabular}}
\end{table}

\begin{table}
\caption{Summary of the change-points detected from (M4).}
\label{table:sim:four}
\centering
\scriptsize{
\begin{tabular}{ c | c | c c c || c c c }
\hline
\hline
& & \multicolumn{3}{c}{$p=50$}
& \multicolumn{3}{c}{$p=100$} \\
$\varrho$ &  & THR & AVG & MAX &
THR & AVG & MAX \\
\hline
\multirow{3}{*}{0.05} & mean & 0.99 &	1.03 &	0.98 &	0.88 &	4.12 &	0.89 \\
 & sd & 0.52 &	1.49 &	0.51 &	0.38 &	2.06 &	0.37 \\
 & $t=100$ & 80 &	35 &	73 &	80 &	87 &	78 \\
\hline
\multirow{3}{*}{0.25} & mean & 1.04 &	1.72 &	0.98 &	1.06 &	4.74 &	1.12 \\
 & sd & 0.24 &	1.58 &	0.51 &	0.34 &	1.98 &	0.67 \\
 & $t=100$ & 91 &	74 &	73 &	93 &	97 &	73 \\
\hline
\multirow{3}{*}{0.5} & mean & 1.14 &	2.1 &	1.03 &	1.09 &	5.56 &	1.09 \\
 & sd & 0.47 &	1.64 &	0.41 &	0.32 &	2.26 &	0.38 \\
 & $t=100$ & 92 &	92 &	74 &	97 &	100 &	62 \\
\hline
\multirow{3}{*}{1} & mean & 1.09 &	2.94 &	1.01 &	1.28 &	0.02 &	1.05 \\
 & sd & 0.32 &	1.9 &	0.41 &	0.73 &	0.14 &	0.39 \\
 & $t=100$ & 94 &	99 &	50 &	97 &	2 &	49 \\
\hline
\end{tabular}}
\end{table}

\section{Detecting change-points in the component processes of S\&P 500}
\label{sec:real}

We further study the performance of the SBS-MVTS algorithm by applying it to
the multivariate time series of \emph{daily closing prices} of the constituents of
the S\&P 500 stock market index.
The period considered is between 1 January 2007 and 31 December 2011,
overlapping with the period of the recent financial crisis. We have chosen only those
$456$ constituents that remained in the index over the entire period;
the resulting time series is of dimensionality $p=456$ and length $T=1260$
(we recall that $d$ is quadratic in $p$ and therefore much larger than
$T$ in this example).

Before presenting the change-point detection results, we briefly mention the rationale behind our approach to this dataset.
As noted in Section \ref{sec:wavper}, the wavelet periodograms computed with Haar wavelets at scale $i=-1$ take the form
$I_{i, t, T} = 2^{-1}(X_{t+1, T}-X_{t, T})^2$ and thus reflect the behaviour of return series,
and these periodograms comprise the input multiplicative sequences to SBS-MVTS.
\citet{mikosch2004} discussed that the ``stylized facts'' observable in financial time series,
such as long range dependence of the absolute returns,
might be artifacts induced by change-points in the second-order structure of the series.
It was further discussed in \citet{piotr2005} where a class of Gaussian LSW time series was shown to embed these stylized facts.

When first applied to the first 100 component processes, the algorithm returns
$t=$ 67, 129, 198, 276, 427, 554, 718, 864, 1044, 1147 as change-points.
We then apply the algorithm to the first 200 processes to obtain
$t=$ 67, 126, 198, 270, 333, 427, 554, 652, 718, 867, 1022, 1086, 1148 as change-points.
Comparing the two sets of detected change-points, it is reassuring to see that those from the former set
also appear to have their counterparts in the latter, as expected, since the latter dataset contains the former.
When applied to the entire $p$-variate time series, the SBS-MVTS algorithm returns the change-points
summarized in Table \ref{table:snp}, which also lists some historical events that occurred
close to some of the detected change-points.

\begin{figure}[htbp]
\centering
\epsfig{file=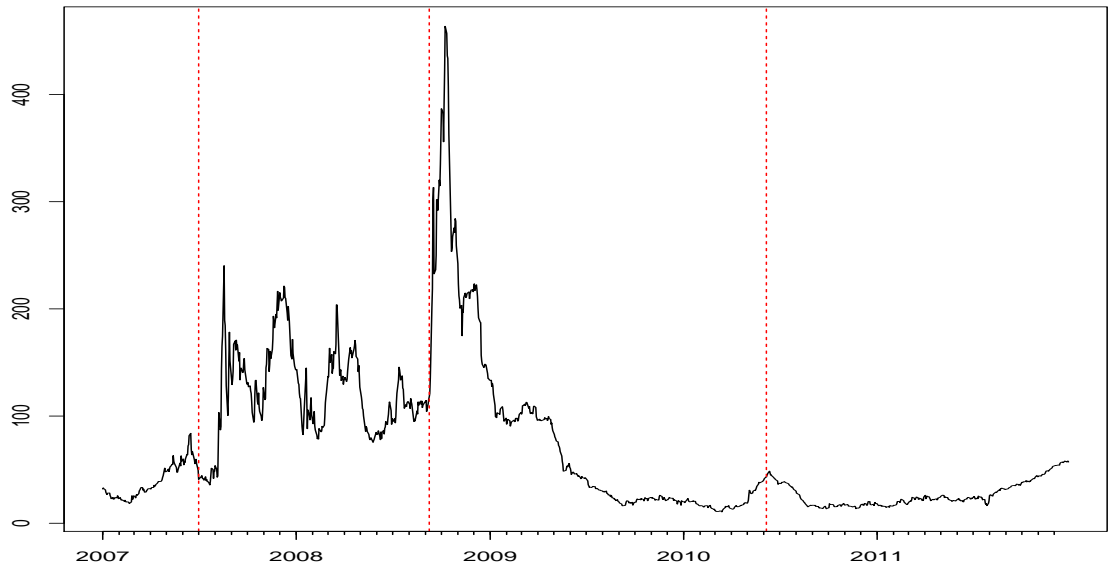, width=.9\linewidth, height=3in}
\caption{
TED spread between 2007 and 2011 with the estimated change-points (vertical lines) marked by $\dagger$ in Table \ref{table:snp}.}
\label{fig:ted}
\end{figure}

The TED spread is the difference between the interest rate at which the US Government is able to borrow
over a three month period (T-bill) and the rate at which banks lend to each other
over the same period (measured by the Libor),
and therefore can serve as an indicator of perceived credit risk in the general economy.
During 2007, the TED spread rapidly increased to around 150--200 basis points (bps),
which coincided with the ``subprime'' mortgage crisis, and in mid-September 2008,
it exceeded 300 bps. In 2010, it returned to its long-term average of 30 bps. However,
it started rising again with the beginning of the European debt crisis, and reached above
45 bps by mid-June. The volatile behaviour of the TED spread during 2007--2011 is reflected in
some of the change-points detected by the SBS-MVTS algorithm as shown in Figure \ref{fig:ted}.

To further check the validity of the detected change-points,
we tested the stationarity of the series within the segments examined at each iteration of the SBS-MVTS algorithm.
The problem of testing stationarity for multivariate time series has not been widely studied;
\citet{suhasini2013} note that only few procedures exist for such a purpose
and those existing ones are not easily applicable to the current dataset with dimensionality as large as $p=456$.

Instead, we chose to examine the stationarity of first few principal component series obtained over each segment.
Various methods have been proposed for testing second-order stationarity of univariate time series
and among them, the multiple testing procedure proposed in \citet{nason2013} is available in the format of an R package.
However, since its test statistics are close to ours except that they are computed at the locations which are power of two,
we concluded that performing this procedure would not be suitable for our purpose.

Alternatively, we adopted the stationarity test proposed in \citet{dwivedi2011} (R code is available on \url{http://www.stat.tamu.edu/~suhasini/Rcode.html}),
which tests whether the correlations between the discrete Fourier transforms of the series are close to zero.
We applied the testing procedure to each segment examined as the SBS-MVTS algorithm proceeded.
That is, since change-points were detected in the order $550$, $\{426, 1148\}$, $\{199, 1017\}$, $\{126, 274, 711, 1088\}$, $\{66, 864\}$
(those detected at the same ``level'' were grouped together),
we investigated the segments $[1, 1260]$, $[1, 549]$, $[550, 1260]$, $[1, 425]$, $[426, 549]$ and so forth.
Within each segment $[s, e]$, principal component analysis was performed on $\mathbf{X}_t$,
producing two factor series as the first two principal components.
As these factors often exhibited high autocorrelations (which might falsely lead to rejecting the null hypothesis),
we fit an AR(1) process to each factor and tested the stationarity of these residual series.

Furthermore, we checked whether the resulting residuals behaved like Gaussian white noise.
It may be expected that if $\mathbf{X}_t$ is stationary within $t \in [s, e]$,
the residuals behave like Gaussian white noise under our LSW model,
whereas if its second-order structure undergoes a change,
departure from Gaussianity is observable from the distribution of the residuals.
To do so, we adopted the normality tests which were implemented on R (packages \texttt{tseries} and \texttt{nortest}),
namely Lilliefors, Anderson-Darling, Pearson, Shapiro-Francia and Jarque-Bera tests.
While failing to reject the null hypothesis via these tests do not guarantee that the residual series follows a normal distribution,
they can serve as an indicator that certain moments and quantiles of the residuals behave like those of Gaussian random variables.

Adopting the Bonferroni correction as in \citet{nason2013}, we rejected the null hypothesis of stationarity or normality
when the corresponding p-value was smaller than $\alpha^* = 0.05/23 = 0.00212$
(dependence in the test statistics was not taken into account).
For most of the segments containing any change-points, the p-values were smaller than $\alpha^*$
for at least one of the factors,
except for $[119, 425]$ (for normality tests) and $[1017, 1147]$ (for both tests).
On the other hand, p-values were generally greater than $\alpha^*$ over the segments
which did not contain any change-point,
indicating that the residuals over these segments behaved similarly as Gaussian white noise.
Some segments, such as $[1, 65]$, both of the null hypotheses were rejected
which implies that further change-points could have been detected
but the restriction imposed on change-point dispersion in the SBS algorithm
prevented them from being detected.

Overall, the findings support the use of the SBS-MVTS methodology in this case study.

\begin{table}
\caption{Summary of the change-points detected from the component processes of S\&P 500;
refer to the TED spread in Figure \ref{fig:ted} for the change-points marked by $\dagger$.}
\label{table:snp}
\centering
\small{
\begin{tabular}{| c | c | l |}
\hline
\hline
t & date & historical event \\
\hline
66 & 2007/04/09 & \\
126 & 2007/07/03 & TED spread$^\dagger$ \\
199 & 2007/10/16 & US stock market peaked in October 2007. \\
274 & 2008/02/04 &  \\
426 & 2008/09/10 & TED spread$^\dagger$ \\
\multirow{2}{*}{550} & \multirow{2}{*}{2009/03/10} & The Dow Jones average index reached a trough of around 6600 by March 2009; \\
& & identified by the New York Times as the ``nadir of the crisis". \\
711 & 2009/10/27 & \\
864 & 2010/06/08 & TED spread$^\dagger$ \\
1017 & 2011/01/13 & \\
1088 & 2011/04/27 & \\
1148 & 2011/07/22 & Global stock markets fell due to fears of contagion of the European sovereign debt crisis. \\
\hline
\end{tabular}}
\end{table}

\vspace{10pt}

{\bf Acknowledgements}

We would like to thank the Editor, Associate Editor and two Referees for very helpful comments which
led to a substantial improvement of this manuscript.

\vspace{10pt}

\appendix
\section{Proof of Theorem \ref{thm:one}}
\label{app:thm:one}

We first prove a set of lemmas that are essential in proving Theorem \ref{thm:one}
for a single multiplicative sequence following model (\ref{generic:one}).
Note that when $d=1$, the algorithm returns identical change-points
no matter if $\ty^{\mbox{\scriptsize thr}}_t$ or the raw CUSUM statistic $\cY^{(1)}_{s, t, e}$
are used.
In this section, the superscripts are suppressed where there is no confusion.
Define $\bbY_{s, b, e}$ as
\begin{eqnarray}
\bbY_{s, b, e}
=\left\vert\sqrt{\frac{e-b}{n \cdot (b-s+1)}} \sum_{t=s}^b \ytt-
\sqrt{\frac{b-s+1}{n\cdot(e-b)}}\sum_{t=b+1}^e \ytt\right\vert.
\nonumber
\end{eqnarray}
for $n=e-s+1$, and $\bbS_{s, b, e}$ is defined similarly with $\sigt$ replacing $\ytt$.
Further, let $\eta_1 < \eta_2 < \ldots < \eta_N$ be the change-points in $\sigt$
(with the convention of $\eta_0=0$ and $\eta_{N+1}=T-1$).
In what follows, $c_i, \ i=1, 2, \ldots$ are used to denote specific positive constants
and $C, C'$ to denote generic ones.

Let $s$ and $e$ denote the ``start'' and the ``end'' of a segment
to be examined at some stage of the algorithm.
Further, we assume that $s$ and $e$ satisfy
\begin{eqnarray*}
\eta_{q_1} \le s < \eta_{q_1+1} < \ldots < \eta_{q_2} < e \le \eta_{q_2+1}
\end{eqnarray*}
for $0 \le q_1 < q_2 \le N$.
In Lemmas \ref{lem:one}--\ref{lem:five}, we impose at least one of following conditions:
\begin{eqnarray}
s< \eta_{q_1+q}-c_1\delt < \eta_{q_1+q}+c_1\delt < e \mbox{ for some } 1 \le q \le q_2-q_1,
\label{lem:cond:one} \\
\{(\eta_{q_1+1}-s)\wedge(s-\eta_{q_1})\} \vee \{(\eta_{q_2+1}-e)\wedge(e-\eta_{q_2})\} \le c_2\ept,
\label{lem:cond:two}
\end{eqnarray}
where $\wedge$ and $\vee$ are the minimum and maximum operators.
We later show that under (A1)--(A4), both conditions (\ref{lem:cond:one}) and (\ref{lem:cond:two}) hold
throughout the algorithm for all those segments which contain change-points still to be detected.
Finally, throughout the following proofs,
$\delt$ and $\ept$ are as assumed in Theorem \ref{thm:one}
along with the threshold $\thr$ and other quantities involved in their definitions, i.e. $\theta, \vartheta, \kappa, \gamma$ and $\omega$.

\vspace{10pt}

\begin{lem}
\label{lem:one}
Let $s$ and $e$ satisfy (\ref{lem:cond:one}). Then there exists $1 \le q^* \le q_2-q_1$ such that
\begin{eqnarray}
\left|\bbS_{s, \eta_{q_1+q^*}, e}\right| \ge c_3\frac{\delt}{\sqrt{T}}.
\label{lem:one:eq:one}
\end{eqnarray}
\end{lem}
\noindent \textbf{Proof.}
When there exists a single change-point in $\sig(z)$ over $(s, e)$,
we have $q^*=1$ and thus use the constancy of $\sig(z)$ to the left and the right of $\eta_{q_1+q^*}$ to show that
\begin{eqnarray*}
\left\vert\bbS_{s, \eta_{q_1+q^*}, e}\right\vert
=\sqrt{\frac{(\eta_{q_1+q^*}-s+1)(e-\eta_{q_1+q^*})}{n}}
\left\vert
\sig\left(\frac{\eta_{q_1+q^*}+1}{T}\right)-\sig\left(\frac{\eta_{q_1+q^*}}{T}\right)\right\vert,
\end{eqnarray*}
which is bounded from below by $\sig_*c_1\delt/\sqrt{T}$ from (A1) and (A3).
In the case of multiple change-points, we remark that for any $q$ satisfying (\ref{lem:cond:one}),
there exists at least one $q^*$ for which
\begin{eqnarray}
\left\vert\frac{1}{\eta_{q_1+q^*}-s+1}\sum_{t=s}^{\eta_{q_1+q^*}}\sig\left(\frac{t}{T}\right)-
\frac{1}{e-\eta_{q_1+q^*}}\sum_{t=\eta_{q_1+q^*}+1}^{e}\sig\left(\frac{t}{T}\right)\right\vert
\label{lem:one:eq:two}
\end{eqnarray}
is bounded away from zero under (A3).
Therefore, the same arguments apply as in the case of a
 single change-point and (\ref{lem:one:eq:one}) follows. \hfill$\square$

\vspace{10pt}

\begin{lem}
\label{lem:two}
Suppose (\ref{lem:cond:one}) holds.
Then there exists $c_0\in(0, \infty)$ such that for $b$ satisfying
$|\eta_{q_1+q}-b| \ge c_0\ept$ and $\bbS_{s, b, e} < \bbS_{s, \eta_{q_1+q}, e}$ for some $q$,
we have $\bbS_{s, \eta_{q_1+q}, e} \ge \bbS_{s, b, e}+C\ept/\sqrt{T}$.
\end{lem}
\noindent \textbf{Proof.}
Without loss of generality, let $\eta\equiv\eta_{q_1+q}<b$.
Then we have
\begin{eqnarray*}
\bbS_{s, b, e}
=\frac{\sqrt{\eta-s+1}\sqrt{e-b}}{\sqrt{e-\eta}\sqrt{b-s+1}}\bbS_{s, \eta, e},
\end{eqnarray*}
and therefore using the Taylor expansion and Lemma \ref{lem:one},
\begin{eqnarray*}
&& \bbS_{s, \eta, e}-\bbS_{s, b, e}
=
\left(1-\frac{\sqrt{\eta-s+1}\sqrt{e-b}}{\sqrt{e-\eta}\sqrt{b-s+1}}\right) \cdot \bbS_{s, \eta, e}
=
\frac{\sqrt{1+\frac{b-\eta}{\eta-s+1}}-\sqrt{1-\frac{b-\eta}{e-\eta}}}
{\sqrt{1+\frac{b-\eta}{\eta-s+1}}} \cdot \bbS_{s, \eta, e}
\\
&\ge&
\frac{\left(1+\frac{c_0\ept}{2c_1\delt}\right)-\left(1-\frac{c_0\ept}{2c_1\delt}\right)+o\left(\frac{c_0\ept}{n}\right)}
{\sqrt{2}}\cdot\bbS_{s, \eta, e}
\ge
\frac{c_0\ept}{c_1\sqrt{2}\delt}\cdot c_3\frac{\delt}{\sqrt{T}} = C\frac{\ept}{\sqrt{T}}.
\qquad \qquad \qquad \quad \square
\end{eqnarray*}

\vspace{10pt}

\begin{lem}
\label{lem:three}
Define
\begin{eqnarray}
\cD=\left\{1 \le s<b<e\le T; \ n \equiv e-s+1 \ge \delt \mbox{ and }
\max\left(\frac{b-s+1}{n}, \frac{e-b}{n}\right) \le c_*\right\}
\nonumber
\end{eqnarray}
for the same $c_*$ as that used in (\ref{assum:eq}).
Then as $T \to \infty$,
\begin{eqnarray}
\p\left(\max_{(s, b, e)\in\cD}\left|\bbY_{s, b, e}-\bbS_{s, b, e}\right| > \log\,T\right) \to 0.
\label{lem:three:eq}
\end{eqnarray}
\end{lem}
\noindent \textbf{Proof.}
We first study the probability of the following event
\begin{eqnarray}
\frac{1}{\sqrt{n}}\left\vert\sumse c_t\cdot\sigt(\zstt-1)\right\vert>\log\,T,
\label{lem:three:one}
\end{eqnarray}
where $c_t=\sqrt{(e-b)/(b-s+1)}$ for $s \le t \le b$ and $c_t=-\sqrt{(b-s+1)/(e-b)}$ for $b+1 \le t \le e$.
Note that from the definition of $\cD$, we have
$|c_t| \le c^* \equiv \sqrt{\frac{c_*}{1-c_*}} < \infty$.
Let $\{U_i\}_{i=1}^{n}$ denote i.i.d. standard normal variables,
$\bV=(v_{i, j})_{i, j=1}^{n}$ with $v_{i, j}=\corr\left(Z_{i, T}, Z_{j, T}\right)$,
and $\bW=(w_{i, j})_{i, j=1}^{n}$ be a diagonal matrix with $w_{i, i}=c_t\cdot\sigt$ where $t=i+s-1$.
By standard results (see e.g. \citet{jk1970}, page 151),
the probability of the event (\ref{lem:three:one}) equals
$\p(n^{-1/2} |\sum_{i=1}^{n} \lam_i(U_i^2-1)| >\log\,T)$,
where $\lam_i$ are eigenvalues of the matrix $\mathbf{VW}$.
Due to the Gaussianity of $U_i$, it follows that $\lam_i(U_i^2-1)$ satisfy the Cram\'{e}r's condition,
i.e., there exists a constant $C>0$ such that
\begin{eqnarray*}
\E\left\vert\lam_i(U_i^2-1)\right\vert^k\le C^{k-2}k!\cdot
\E\left\vert\lam_i(U_i^2-1)\right\vert^2, \ k=3, 4, \ldots.
\end{eqnarray*}
Therefore we can apply the Bernstein inequality \citep{bosq1998} and obtain
\begin{eqnarray}
\p\left(\left\vert\sumse c_t\cdot\sigt(\zstt-1)\right\vert > \sqrt{n}\log\,T\right) \le
2\exp\left(-\frac{n\log^2\,T}{4\sum_{i=1}^{n}\lam_i^2+2\max_i|\lam_i|C\sqrt{n}\log\,T}\right).
\nonumber
\end{eqnarray}
It holds that
$\sum_{i=1}^{n}\lam_i^2=\tr(\mathbf{VW})^2 \le c^{*2}\max_z\sig^2(z)n\phi_{\infty}^2$.
We also note that $\max_i|\lam_i| \le c^*\max_z\sig(z)\Vert\bV\Vert_2$,
where $\Vert\cdot\Vert_2$ denotes the spectral norm of a matrix,
and that $\Vert\bV\Vert_2 \le \phi^1_{\infty}$.
Then, the probability in (\ref{lem:three:eq}) is bounded from above by
\begin{eqnarray*}
\sum_{(s, b, e)\in\cD} 2\exp\left(-\frac{n\log^2\,T}
{4c^{*2}\max_z\sig^2(z)n\phi_{\infty}^2+2c^*\max_z\sig(z)\sqrt{n}\log\,T\phi^1_{\infty}}\right)
\le 2T^3\exp\left(-C'\log^2\,T\right)
\end{eqnarray*}
which converges to 0, since $\phi_{\infty}^1 < \infty$ from (A2), $n \ge \delt > \log\,T$ and
$c^* < \infty$.
\hfill$\square$

\vspace{10pt}

\begin{lem}
\label{lem:four}
Under (\ref{lem:cond:one}) and (\ref{lem:cond:two}),
define an interval $\cD_{s, e}=\{t\in(s, e); \ \max\{(t-s+1)/n, (e-t)/n\} \le c_*\} \subset [s, e]$.
Then there exists $1 \le q^* \le q_2-q_1$ such that $\eta_{q_1+q^*} \in \cD_{s, e}$ and
$|\heta-\eta_{q_1+q^*}|<c_0\ept$ for $\heta = \arg\max_{t\in\cD_{s, e}}|\bbY_{s, t, e}|$.
\end{lem}
\noindent \textbf{Proof.}
The following proof is an adaptation of the proof of Theorem 3.1 in \citet{piotr2013} to non-Gaussian and non-i.i.d. noise.

We note that the model (\ref{generic:one}) can be re-written as
$$\ytt = \sigt+\sigt(\zstt-1), \quad t=0, \ldots, T-1,$$
which in turn can be regarded as a generic additive model $y_t = f_t + \vep_t$ with a piecewise-constant signal $f_t$
by setting $y_t = \ytt$, $f_t = \sigt$ and $\vep_t = \sigt(\zstt-1)$.

On a given segment $[s, e]$, detecting a change-point is equivalent to
fitting the best step function (i.e. a piecewise constant function with one change-point)
$\wh{f}_t$ which minimizes $\sumse(y_t-g_t)^2$
among all step functions $g_t$ defined on $[s, e]$.
Let $f^0_t$ denote the best step function approximation to $f_t$ with its change-point located within $\cD_{s, e}$,
i.e. any $g_t$ which has its change-point in $\cD_{s, e}$ and minimizes $\sumse(f_t-g_t)^2$
($f^0_t$ may or may not be unique).
Under (A1) and (\ref{lem:cond:one})--(\ref{lem:cond:two}), Lemmas 2.2--2.3 in \citet{venkatraman1993} imply that
the single change-point in $f^0_t$ coincides with one of any undetected change-points of $f_t$ in $\cD_{s, e}$,
and we denote such a change-point by $\eta$.

Let us assume that $\wh{f}_t$ has a change-point at $t=\heta$ and it satisfies $|\heta-\eta|=c_0\ept$.
Then if we show
\begin{eqnarray}
\sumse(y_t-f^0_t)^2 - \sumse(y_t-\wh{f}_t)^2 < 0,
\label{lem:four:wts}
\end{eqnarray}
it would prove that $\heta$ must be within the distance less than $c_0\ept$ from $\eta$.
Expanding the left-hand side of (\ref{lem:four:wts}), we obtain
\begin{eqnarray*}
\sumse(\vep_t+f_t-f^0_t)^2 - \sumse(\vep_t+f_t-\wh{f}_t)^2
= 2\sumse\vep_t(\wh{f}_t-f^0_t) + \sumse \{(f_t-f^0_t)^2 - (f_t-\wh{f}_t)^2\}
\equiv I + II.
\end{eqnarray*}
From the definition of $f^0_t$, it is clear that $II<0$. Let ${\mathcal F}$ be the set of vectors
that are initially constant and positive, then contain a change-point, following which are
constant and negative; moreover, they sum to zero and to one when squared.
Let $\bar{f}$ be the mean of $f_t$ on $t\in[s, e]$, and the vector $\psi^0 \in {\mathcal F}$ satisfy
$f^0_t = \bar{f} + \langle f, \psi^0 \rangle\psi^0$.
Then we have
\begin{eqnarray}
\sumse(f_t-f^0_t)^2 &=& \sumse(f_t-\bar{f})^2 -
2\langle f, \psi^0 \rangle\sumse(f_t-\bar{f})\psi^0
+ \langle f, \psi^0 \rangle^2\sumse (\psi^0)^2
\nonumber \\
&=& \sumse(f_t-\bar{f})^2 - \langle f, \psi^0 \rangle^2.
\label{lem:four:eq:one}
\end{eqnarray}
Let a step function $\tilde{f}_t$ be chosen so as to minimize $\sumse(f_t-g_t)^2$ under the constraint that
$g_t$ shares the same change-point as $\wh{f}_t$.
Then we have
\begin{eqnarray}
\sumse(f_t-\tilde{f}_t)^2 \le \sumse(f_t-\wh{f}_t)^2
\label{lem:four:eq:two}
\end{eqnarray}
Representing $\tilde{f}_t = \bar{f} + \langle f, \tilde{\psi} \rangle\tilde{\psi}$ for another vector
$\tilde{\psi} \in {\mathcal F}$ and using (\ref{lem:four:eq:one}) and (\ref{lem:four:eq:two}),
\begin{eqnarray*}
&& \sumse\{(f_t-f^0_t)^2 - (f_t-\wh{f}_t)^2\} \le \sumse\{(f_t-f^0_t)^2 - (f_t-\tilde{f}_t)^2\}
= \langle f, \tilde{\psi} \rangle^2 - \langle f, \psi^0 \rangle^2
\\
&=& (|\langle f, \tilde{\psi} \rangle| - |\langle f, \psi^0 \rangle|)
(|\langle f, \tilde{\psi} \rangle| + |\langle f, \psi^0 \rangle|)
\le (|\langle f, \tilde{\psi} \rangle| - |\langle f, \psi^0 \rangle|)
|\langle f, \psi^0 \rangle|.
\end{eqnarray*}
Since $|\langle f, \psi^0 \rangle| = \bbS_{s, \eta, e}$ and $|\langle f, \tilde{\psi} \rangle| = \bbS_{s, \heta, e}$
with the distance between $\eta$ and $\heta$ being at least $c_0\ept$,
the above is bounded from above by $-C\delt/\sqrt{T} \cdot \ept/\sqrt{T} = -C\ept\delt/T$
from Lemmas \ref{lem:one}--\ref{lem:two}.

Turning to $I$, we can decompose the term as
\begin{eqnarray*}
\sumse\vep_t(\wh{f}_t-f^0_t) = \sumse\vep_t(\wh{f}_t-\tilde{f}_t) + \sumse\vep_t(\tilde{f}_t-f^0_t),
\end{eqnarray*}
and each of the two sums are split into sub-sums computed over the intervals of constancy of
$\wh{f}_t-\tilde{f}_t$ and $\tilde{f}_t-f^0_t$, respectively.
Assume $\heta \ge \eta$ without loss of generality, we have
\begin{eqnarray*}
\sumse\vep_t(\tilde{f}_t-f^0_t) =
\left(\sum_{t=s}^{\eta}+\sum_{t=\eta+1}^{\heta}+\sum_{t=\heta+1}^{e}\right)\vep_t(\tilde{f}_t-f^0_t)
\equiv III + IV + V.
\end{eqnarray*}
As $T \to \infty$, we have with probability tending to 1 (Lemma \ref{lem:three})
\begin{eqnarray*}
|III| &=& \left\vert\frac{1}{\sqrt{\eta-s+1}}\sum_{t=s}^{\eta}\vep_t\right\vert \cdot \sqrt{\eta-s+1} \cdot
\left\vert \frac{1}{\heta-s+1}\sum_{t=s}^{\heta}f_i - \frac{1}{\eta-s+1}\sum_{t=s}^{\eta}f_i \right\vert
\\
&\le& C\log\,T\sqrt{\eta-s+1}\cdot\frac{c_0\ept}{\heta-s+1} \le C'\ept\delt^{-1/2}\log\,T.
\end{eqnarray*}
$|V|$ is of the same order as $|III|$ and similarly $|IV|$ is bounded by $C\ept^{1/2}\log\,T$.

As for $\sumse\vep_t(\wh{f}_t-\tilde{f}_t)$, we have
\begin{eqnarray}
\sumse\vep_t(\wh{f}_t-\tilde{f}_t) =
\left(\sum_{t=s}^{\heta}+\sum_{t=\heta+1}^{e}\right)\vep_t(\wh{f}_t-\tilde{f}_t)
\equiv VI+VII.
\nonumber
\end{eqnarray}
Note that $VI$ and $VII$ are of the same order, and with probability converging to 1 as $T \to \infty$,
\begin{eqnarray*}
|VI| = \frac{1}{\heta-s+1} \left(\sum_{t=s}^{\heta}\vep_t\right)^2 = \log^2\,T.
\end{eqnarray*}
Putting together all the above requirements, as long as
\begin{eqnarray}
\frac{\ept\delt}{T} > (\ept\delt^{-1/2}\log\,T) \vee (\ept^{1/2}\log\,T) \vee (\log^2\,T),
\label{lem:four:eq:three}
\end{eqnarray}
the dominance of the term $II$ over $I$ holds and thus we prove the lemma.

From (\ref{lem:four:eq:three}), it is derived that $\Theta>2/3$ and $\ept>\delt^{-2} \cdot T^2\log^2\,T$,
i.e. letting $\ept=\max(T^\theta, \log^{2+\vartheta}\,T)$, it is sufficient to have $\theta \ge 2-2\Theta$ and $\vartheta>0$.
Also the proof of Lemmas \ref{lem:five}--\ref{lem:six} require
$\delt^{-1}T\log\,T \ll \sqrt{\ept} \ll \thr \ll \delt(T\log\,T)^{-1/2}$,
which is satisfied by $\theta=2-2\Theta$ and $\thr=\kappa T^\gamma$ with any $\gamma\in(1-\Theta, \Theta-1/2)$ when $\Theta \in (3/4, 1)$,
and by $\thr = \kappa \log^{1+\omega}\,T$ with any $\omega>\vartheta/2$ when $\Theta = 1$.
\hfill $\square$

\vspace{10pt}

\begin{lem}
\label{lem:five}
Under (\ref{lem:cond:one}) and (\ref{lem:cond:two}),
we have
\begin{eqnarray}
\p\left(\left|\bbY_{s, b, e}\right|<\thr \cdot n^{-1}\sumse\ytt\right) \to 0
\label{lem:five:eq}
\end{eqnarray}
for $b=\arg\max_{t\in\cD_{s, e}}|\bbY_{s, t, e}|$, as $T \to \infty$.
\end{lem}
\noindent \textbf{Proof.}
Define the two events $\cA$ and $\cB$ as
\begin{eqnarray}
\cA=\left\{
\left|\bbY_{s, b, e}\right|<\thr\cdot\frac{1}{n}\sumse\ytt \right\} \mbox{ \ and \ }
\cB=\left\{
\frac{1}{n}\left\vert
\sumse\ytt-\sumse\sigt\right\vert < \bar{\sig}\equiv\frac{1}{2n}\sumse\sigt \right\}.
\nonumber
\end{eqnarray}
We can show that $\p(\cB) \to 1$ as $T \to \infty$ using the Bernstein inequality as in the proof of Lemma \ref{lem:three}
and that the convergence rate is faster than that of (\ref{lem:three:eq}).
Hence $\p(n^{-1}\sumse\ytt \in (\bar{\sig}/2, 3\bar{\sig}/2)) \to 1$.
Since the probability in (\ref{lem:five:eq}) is bounded from above by $\p(\cA\cap\cB)+\p(\cB^c)$,
we only need to show that $\p(\cA\cap\cB) \to 0$.
From Lemma \ref{lem:four}, we have some $\eta\equiv\eta_{q_1+q}$ satisfying $|b-\eta|<c_0\ept$.
Without loss of generality, let $\eta<b$ and define
$\sig_1\equiv\sig\left(\frac{\eta}{T}\right) \ne \sig\left(\frac{\eta+1}{T}\right)\equiv\sig_2$.
From Lemma \ref{lem:three}, (\ref{lem:cond:one})--(\ref{lem:cond:two}) and (A1),
the following holds with probability tending to 1 as $\gamma < \Theta-1/2$:
\begin{eqnarray*}
\left\vert\bbY_{s, b, e}\right\vert &\ge& \left\vert\bbS_{s, b, e}\right\vert-\log\,T
\\
&=&
\sqrt{\frac{(b-s+1)(e-b)}{n}}\left\vert \frac{\sig_1(\eta-s+1)+\sig_2(b-\eta)}{b-s+1}
-\sig_2 \right\vert-\log\,T
\\
&\ge& \sqrt{\frac{e-b}{n(b-s+1)}}\cdot\sig_*(\eta-s+1) - \log\,T
\ge \sqrt{\frac{1-c_*}{nc_*}}\cdot\sig_*(\eta-s+1) - \log\,T \\
&\ge& \frac{C\delt}{c^*\sqrt{T}} - \log\,T > \thr\cdot\frac{3\bar{\sig}}{2}.
\end{eqnarray*}
\hfill$\square$

\vspace{10pt}

\begin{lem}
\label{lem:six}
For some positive constants $C, \ C'$, let $s$, $e$ satisfy either
\begin{itemize}
\item[(i)] $\exists\, 1 \le q \le N$ such that $s \le \eta_q \le e$ and
$(\eta_q-s+1) \wedge (e-\eta_q) \le C\ept$\, or
\item[(ii)] $\exists\, 1 \le q \le N$ such that $s \le \eta_q < \eta_{q+1} \le e$ and
$(\eta_q-s+1) \vee (e-\eta_{q+1}) \le C'\ept$.
\end{itemize}
Then as $T \to \infty$,
\begin{eqnarray}
\p\left(
\left|\bbY_{s, b, e}\right|>\thr \cdot n^{-1}\sumse\ytt\right)\to 0
\label{lem:six:eq}
\end{eqnarray}
for $b=\arg\max_{t\in\cD_{s, e}}|\bbY_{s, t, e}|$.
\end{lem}
\noindent \textbf{Proof.}
First we assume (i).
We define the event $\cA '$ as
$\cA '=\{|\bbY_{s, b, e}| > \thr\cdot n^{-1}\sumse\ytt\}$
and adopt the event $\cB$ from the proof of Lemma \ref{lem:five}.
Since $\p(\cB) \to 1$,
the probability in (\ref{lem:six:eq}) is bounded from above by $\p(\cA '\cap\cB)+\p(\cB^c)$ and
it only remains to show $\p(\cA '\cap\cB) \to 0$.
Assuming $\eta_q-s+1 \le C\ept$ leads to $b>\eta_q\equiv\eta$,
and using the same notation as in Lemma \ref{lem:five} we have
\begin{eqnarray*}
\left\vert\bbY_{s, b, e}\right\vert &\le& \left\vert\bbS_{s, b, e}\right\vert+\log\,T
\\
&\le&
\sqrt{\frac{(b-s+1)(e-b)}{n}}\left\vert
\frac{\sig_1(\eta-s+1)+\sig_2(b-\eta)}{b-s+1}-\sig_2\right\vert+\log\,T
\\
&\le&
\sqrt{\frac{e-b}{n(b-s+1)}}\cdot 2\sig^*(\eta-s+1) + \log\,T
\le
\sqrt{\frac{e-\eta}{n(\eta-s+1)}}\cdot 2\sig^*(\eta-s+1) + \log\,T
\\
&\le& 2\sig^*\sqrt{C\ept} + \log\,T  < \thr\cdot\frac{\bar{\sig}}{2}.
\end{eqnarray*}
The proof in the case of (ii) takes similar arguments and thus Lemma \ref{lem:six} follows.
\hfill$\square$
\vspace{2mm}

When applying the algorithm to a single sequence with $N$ change-points,
Lemmas \ref{lem:one}--\ref{lem:six} shows the consistency of the algorithm as follows.
At the start of the binary segmentation algorithm, we have $s=0$ and $e=T-1$,
and thus all the conditions required by Lemma \ref{lem:five} are met.
Then the algorithm detects and locates a change-point which is within the distance of $c_0\ept$
from a true change-point (Lemma \ref{lem:four}) such that any segments defined by the detected change-points also satisfy the conditions in Lemma \ref{lem:five},
from the assumptions on the spread of $\eta_q, \ q=1, \ldots, N$ in (A1).
The algorithm iteratively proceeds in this manner until all the $N$ change-points are detected,
and since thus-determined segments meet either of the two conditions in Lemma \ref{lem:six},
change-point detection is completed.

Now we turn our attention to the case of $d>1$ sequences and prove Theorem \ref{thm:one}.
When necessary to highlight the dependence of $\yktt$ on $k$
in deriving $\bbS_{s, t, e}$ and $\bbY_{s, t, e}$,
we use the notations $\bbS^{(k)}_{s, t, e}$ and $\bbY^{(k)}_{s, t, e}$.
The index set $\{1, \ldots, d\}$ is denoted by $\cK$.
From Lemma \ref{lem:three}, we have $\max_k\max_{(s, t, e)\in\cD}|\bbY^{(k)}_{s, t, e} - \bbS^{(k)}_{s, t, e}| \le \log\,T$
with the probability bounded from below by
$1-CdT^3\exp\left(-C'\log^2\,T\right) \to 1$ under (A4).
Therefore, the following arguments are made conditional on this event.

Let $\cK_{s, e} \subset \cK$ denote the index set corresponding to those $\yktt$
with at least one change-points in $\sigkt$ on $t\in(s, e)$.
Lemma \ref{lem:six} shows that $\cY^{(k)}_{s, t, e}, \ k\in\cK\setminus\cK_{s, e}$
do not pass the thresholding at any $t\in(s, e)$, i.e.
$\cI^{(k)}_{s, t, e}=\bbI(\cY^{(k)}_{s, t, e} > \thr)=0$ for all $t\in(s, e)$.
On the other hand, Lemma \ref{lem:five} indicates that all $\cY^{(k)}_{s, t, e}, \ k\in\cK_{s, e}$
survive after thresholding in the sense that
$\cI^{(k)}_{s, t, e}=1$ over the intervals around the true change-points.
Besides, in \citet{venkatraman1993}, each $\bbS^{(k)}_{s, t, e}$ is shown to be of the functional form
$g^{(k)}(x)=(x(1-x))^{-1/2}(\alpha^{(k)}_x x+\beta^{(k)}_x)$ for $x=(t-s+1)/n\in(0, 1)$,
where $\alpha^{(k)}_x$ and $\beta^{(k)}_x$ are determined by the magnitude of
the jumps at the change-points of $\sigkt$ as well as their locations,
and constant between any two adjacent change-points.
Note that scaling of $\bbS^{(k)}_{s, t, e}$ by $n^{-1}\sumse\yktt$
scales the values of $\alpha_x$ and $\beta_x$ only,
and does not change the shape of $g^{(k)}(x)$.
Each function $g^{(k)}(x)$
\begin{itemize}
\item[(a)] is either monotonic or decreasing and then increasing on any interval
defined by two adjacent change-points of $\sigkt$, and
\item[(b)] achieves the maximum at one of the change-points of $\sigkt$ in $(s, e)$,
\end{itemize}
see Lemma 2.2 of \citet{venkatraman1993}.
Since the point-wise summation of $g^{(k)}(\cdot)$ over $k\in\cK_{s, e}$
takes the functional form
$g(x) = (x(1-x))^{-1/2}(\alpha_x x+\beta_x)$ which is identical to that of each individual $g^{(k)}(\cdot)$,
it satisfies the above (a)--(b) as well.

Denoting $\bar{Y}_{s, e} = \frac{1}{n}\sum_{u=s}^eY^{(k)}_{u, T}$,
we decompose $\ty^{\mbox{\scriptsize thr}}_t$ as
\begin{eqnarray}
\frac{\ty^{\mbox{\scriptsize thr}}_t}{\sum_{k \in \cK}\cI^{(k)}_{s, t, e}} &=&
\frac{\sum_{k \in \cK} \bar{Y}_{s, e}^{-1} \cdot \bbY^{(k)}_{s, t, e} \cdot \cI^{(k)}_{s, t, e}}{\sum_{k \in \cK}\cI^{(k)}_{s, t, e}}
\nonumber \\
&=&
\frac{\sum_{k \in \cK_{s, e}}\bar{Y}_{s, e}^{-1} \cdot \bbS^{(k)}_{s, t, e}\cdot \cI^{(k)}_{s, t, e}}{|\cK_{s, e}|}
+
\frac{\sum_{k \in \cK_{s, e}}\bar{Y}_{s, e}^{-1}
\left(\bbY^{(k)}_{s, t, e}-\bbS^{(k)}_{s, t, e}\right)\cdot \cI^{(k)}_{s, t, e}}{|\cK_{s, e}|}
= I+II
\nonumber
\end{eqnarray}
where $II\le C\log\,T$ (Lemma \ref{lem:three}).
Note that we can construct an additive model $y_t = f_t + \vep_t$ over $t\in[s, e]$
as the one introduced in Lemma \ref{lem:four},
such that the CUSUM statistic of the piecewise constant signal $f_t$
(i.e. $\bbS_{s, t, e}$ with $f_t$ replacing $\sigt$) is equal to
$|\cK_{s, e}|^{-1}\sum_{k \in \cK_{s, e}}\bar{Y}_{s, e}^{-1} \cdot \bbS^{(k)}_{s, t, e}$.
Since thresholding does not have any impact on the peak formed around the change-points
within the distance of $C \ep_T$,
$I$ is of the same functional form as the CUSUM statistic of $f_t$ in that region
around the change-points.
Therefore from Lemma \ref{lem:four}, $b=\arg\max_{t\in(s, e)}\ty^{\mbox{\scriptsize thr}}_t$
satisfies $|b-\eta_q|<c_0\ept$ for some $q=1, \ldots, N$.

The SBS algorithm continues the change-point detection procedure on the segments
defined by previously detected change-points,
which satisfy both (\ref{lem:cond:one}) and (\ref{lem:cond:two}) for at least one of $k\in\cK$
until every change-point is detected (as in the case of $d=1$).
Once all $\eta_1, \ldots, \eta_N$ are identified,
each of the resulting segments satisfies either (i) or (ii) in Lemma \ref{lem:six} for all $k\in\cK$
such that the termination condition of the SBS algorithm (Step 2.1) is met.

Note that for any $k\in\cK_{s, e}$, a simple modification of the proof of Lemma \ref{lem:two}
leads to the existence of a positive constant $C$ satisfying
$\bbS^{(k)}_{s, t, e} > \thr$ for $|t-\eta| \le C\ept$,
where $\eta$ is any of the change-points of $\yktt$ within $(s, e)$
at which (\ref{lem:one:eq:two}) is not equal to zero.
Then, the corresponding $\bbY^{(k)}_{s, t, e}$ is also greater than $\thr$
within the distance of $\Delta_T \asymp \ept$ from
$b = \arg\max_{t\in[s, e]} \cY^{(k)}_{s, t, e}$,
and hence the condition on the change-point estimates in Step 2.3 is justified with the choice of $\Delta_T=\lfloor \sqrt{T}/2 \rfloor$.

\section{Multivariate LSW time series}
\label{app:two}

The LSW model enables a time-scale decomposition of a multivariate, possibly high-dimensional process and
thus permits a rigorous estimation of its second-order structure as shown in this section.
The following conditions are imposed on the piecewise constant functions $W^{(j)}_i(z)$ and $\Sigma^{(j, l)}_i(k/T)$,
as well as on the change-points in the second-order structure
for the $p$-variate LSW time series defined in Definition \ref{def:lsw:one}.
\begin{itemize}
\item[(B1)] The following holds for each of the piecewise constant functions $W^{(j)}_i(z)$ and $\Sigma^{(j, l)}_i(z)$
for $j, l=1, \ldots, p; \,i=-1, -2, \ldots$.
\begin{itemize}
\item[$\bullet$] Denoting by $L^{(j)}_i$ the total magnitude of jumps in $\{W^{(j)}_i(z)\}^2$,
the variability of the functions $W^{(j)}_i(z), \ i=-1, -2, \ldots$ is controlled such that
$\sum_{i=-I_T}^{-1} 2^{-i}L^{(j)}_i = O(\log\,T)$ uniformly in $j$ where $I_T = \lfloor \log\,T \rfloor$.
Also, there exists a positive constant $C>0$ such that
$|W^{(j)}_i(z)| \le C2^{i/2}$ uniformly over all $i \le -1$ and $j=1, \ldots, p$.
\item[$\bullet$] Denoting the total magnitude of jumps in $\Sigma^{(j, l)}_i(z)$ by $R^{(j, l)}_i$,
the variability of the functions $\Sigma^{(j, l)}_i(z), \ i=-1, -2, \ldots$ is controlled such that
$\sum_{i=-I_T}^{-1} 2^{-i}R^{(j, l)}_i = O(\log\,T)$ uniformly in $j \ne l$.
\end{itemize}
\item[(B2)] Recall $\bbB$, the set of all change-points in the second-order structure of $\bxtt$ defined in (\ref{lsw:br:set}).
Then $\nu_r\in\bbB, \ r=1, \ldots, N$ satisfy the conditions in (A1) in place of $\eta_q, \ q=1, \ldots, N$.
\end{itemize}

The quantity of interest in modelling a multivariate LSW time series is the Evolutionary Wavelet Spectrum (EWS)
and the Evolutionary Wavelet Cross-spectrum (EWCS), which are defined as
\begin{eqnarray*}
S^{(j)}_i(z) &=& S^{(j, j)}_i(z) = (W^{(j)}_i(z))^2 \mbox{\quad for \ } j=1, \ldots, p, \\
S^{(j, l)}_i(z) &=& W^{(j)}_i(z)W^{(l)}_i(z)\Sigma^{(j, l)}_i(z) \mbox{\quad for \ } j \ne l; \,j, l=1, \ldots, p.
\end{eqnarray*}
To study the connection between EWS and the second-order structure of $\bxtt$,
we adopt the following quantities from \citet{nason2000}:
with the same wavelet system as that used in the definition of $\bxtt$,
we define the autocorrelation wavelets as $\Psi_i(\tau)=\sum_k\psi_{i, k}\psi_{i, k+\tau}$,
the cross-scale autocorrelation wavelets as $\Psi_{i, i'}(\tau)=\sum_k\psi_{i, k}\psi_{i', k+\tau}$, and
the autocorrelation wavelet inner product matrix as $\bA=\left(A_{i, i'}\right)_{i, i'<0}$
with $A_{i, i'}=\sum_\tau\Psi_i(\tau)\Psi_{i'}(\tau) = \sum_\tau\Psi_{i, i'}^2(\tau)>0$.
Then, the \emph{local} autocovariance and cross-covariance functions of $\bxtt$ are defined as
\begin{eqnarray*}
c^{(j)}(z, \tau) =  c^{(j, j)}(z, \tau) &=&
\sumi S^{(j)}_i(z)\Psi_i(\tau) \qquad \mbox{(from \citet{nason2000})},\\
c^{(j, l)}(z, \tau) &=& \sumi S^{(j, l)}_i(z)\Psi_i(\tau) \quad \mbox{(from \citet{jean2010})}.
\end{eqnarray*}
Recalling the definition of time-varying autocovariance and cross-covariance functions,
the functions $c^{(j, l)}(z, \tau)$ and $c^{(j, l)}_T(z, \tau)$ are close to each other in the following sense.

\vspace{10pt}

\begin{prop}
\label{prop:one}
Under (B1)--(B2), $c^{(j, l)}_T(z, \tau)$ converges to $c^{(j, l)}(z, \tau)$ as
\begin{eqnarray}
\frac{1}{T}\sum_{t=0}^{T-1}\left\vert c^{(j, l)}_T(t/T, \tau) - c^{(j, l)}(t/T, \tau) \right\vert = o(1)
\label{asymp:tend}
\end{eqnarray}
for all $j, l=1, \ldots, p$.
\end{prop}
\vspace{2mm}
The proof is provided in Appendix \ref{app:prop:one}.
From (\ref{asymp:tend}), we can see that there exists an asymptotic one-to-one relationship between the EWS (EWCS)
$S^{(j, l)}_i(z)$ and the autocovariance (cross-covariance) functions $c^{(j, l)}_T(z, \tau), \ \tau\in\mathbb{Z}$
for all $j, l=1, \ldots, p$ such that,
if there is a change-point in $c^{(j, l)}_T(z, \tau)$ at some lag $\tau$,
at least one of the corresponding $S^{(j, l)}_i(z), \ i=-1, -2, \ldots$ has a change-point
at the same location $z$, and vice versa.

Furthermore, we can also show an one-to-one correspondence between the EWS (EWCS) and the wavelet periodograms (cross-periodograms).
Let $\beta^{(j, l)}_i(z)$ be a linear transformation of the EWS (EWCS) defined as
$\beta^{(j, l)}_i(z)=\sum_{i'=-\infty}^{-1} S^{(j, l)}_{i'}(z)A_{i, i'}$.
Then, the function $\beta^{(j, l)}_i(z)$ is piecewise constant with its change-points corresponding to those of
$\{S^{(j, l)}_{i'}(z)\}_{i'}$ due to the invertibility of $\bA$.

\vspace{10pt}

\begin{prop}
\label{prop:two}
Under (B1)--(B2), $\E I^{(j, l)}_{i, t, T}$ satisfies
\begin{eqnarray}
\frac{1}{T}\sum_{t=0}^{T-1}\left\vert\E I^{(j, l)}_{i, t, T}-
\beta^{(j, l)}_i\left(\frac{t}{T}\right)\right\vert^2
=2^{-i}O(T^{-1})+b^{(j, l)}_{i, T},
\label{eq:prop:two}
\end{eqnarray}
where $b^{(j, l)}_{i, T}$ depends on the corresponding sequence $\{L^{(j)}_i\}_i$ or $\{R^{(j, l)}_i\}_i$.
\end{prop}
\vspace{2mm}

The proof of (\ref{eq:prop:two}) is a direct modification of that of Proposition 2.1 of \citet{piotr2006a} and thus is omitted.
In summary, from Propositions \ref{prop:one}--\ref{prop:two} and the invertibility of $\bA$,
there exists an asymptotic one-to-one correspondence between
the autocovariance (cross-covariance) functions and the expectations of wavelet periodograms (cross-periodograms)
as noted in Section \ref{sec:link}.
Therefore, any change-points in the autocovariance (cross-covariance) functions are detectable
by examining the corresponding wavelet periodogram (cross-periodogram) sequences.

\subsection{Proof of Theorem \ref{thm:two}}
\label{app:two:two}

From its construction, $\E I^{(j, l)}_{i, t, T}$ is piecewise constant and ``almost'' satisfies (A1) and (A3)
in the sense that, for any change-point $\nu$ in $\beta^{(j, l)}_i(t/T)$,
\begin{itemize}
\item[(a)] $\E I^{(j, l)}_{i, t, T}$ is piecewise constant apart from the intervals $[\nu-K2^{-i}, \nu+K2^{-i}]$
for some $K>0$, where it shows smoother transitions, and
\item[(b)] $\E I^{(j, l)}_{i, t, T}$ has at least one change-point within the intervals $[\nu-K2^{-i}, \nu+K2^{-i}]$,
such that $|\E I^{(j, l)}_{i, t_1, T} - \E I^{(j, l)}_{i, t_2, T}|$ is bounded away from zero
for $t_1=\nu-K2^{-i}-1$ and $t_2=\nu+K2^{-i}+1$.
\end{itemize}
Note that (a) and (b) also hold for $\E\tilde{I}^{(j, l)}_{i, t, T}$ for $j \ne l$ defined as in (\ref{cross}).
To accommodate these features of $\ijitt$ and $\tilde{I}^{(j, l)}_{i, t, T}$,
we propose a modification of the multiplicative model (\ref{generic:one}),
\begin{eqnarray}
\tytt^{(k)}=\sigtt^{(k)}\cdot\tztt^{(k)2}, \quad t=0, \ldots, T-1; \ k=1, \ldots, d.
\label{generic:two}
\end{eqnarray}
The difference between the two models (\ref{generic:one}) and (\ref{generic:two}) comes from
the function $\E\tytt^{(k)} = \sigktt$ which is close to a piecewise constant function $\sigkt$
as $\E I^{(j, l)}_{i, t, T}$ is close to $\beta^{(j, l)}_i(z)$ (see (\ref{eq:prop:two})).

We also adapt the assumptions (A2)--(A4) to the multivariate time series set-up,
and denote their analogues in this setting by (B3)--(B5). The latter assumptions are
imposed on $\ijitt, \ j=1, \ldots, p$ and $\tilde{I}^{(j, l)}_{i, t, T}, \ j \neq l; \, j, l=1, \ldots, p$
at scales $i=-1, \ldots, \is_T$, using the representation of these quantities as in
(\ref{generic:two}) (the notation below refers to that representation).
\begin{itemize}
\item[(B3)]
$\{\tztt^{(k)}\}_{t=0}^{T-1}$ is a sequence of standard normal variables and $\max_k\phi^{(k)1}_{\infty}<\infty$, where
$\phi^{(k)}(\tau)=\sup_{t,T}|\corr(\tztt^{(k)}, \tilde{Z}^{(k)}_{t+\tau, T})|$ and $\phi^{(k)r}_{\infty}=\sum_{\tau}|\phi^{(k)}(\tau)|^r$.
\item[(B4)] There exist positive constants $\sig^*, \sig_*>0$ such that
$\{\max_{k,t,T}\sigkt \vee \max_{k,t,T}\sigktt\} \le \sig^*$,
and given any change-point $\eta_q$ in $\sigkt$, we have
$|\sig^{(k)}((\eta_q+1)/T)-\sig^{(k)}(\eta_q/T)| > \sig_*$ uniformly for all $k$.
\item[(B5)] $p$ and $T$ satisfy $p^2 \cdot T^{-\log\,T} \to 0$.
\end{itemize}

The following proposition shows that applying the SBS algorithm to $\tytt^{(k)}$
instead of $\yktt$ also leads to consistent change-point estimates $\teta_q, \ q=1, \ldots, \tilde{N}$.

\vspace{10pt}

\begin{prop}
\label{prop:three}
Under (A1), (A4) and (B3)--(B4), letting $\Delta_T \asymp \ept$, we have that
$\teta_q, \ q=1, \ldots, \tilde{N}$ satisfy
\begin{eqnarray}
\p\left\{\tilde{N}=N; \,|\teta_q-\eta_q| < C_3\ept \mbox{ for } q=1, \ldots, N\right\} \to 1
\nonumber
\end{eqnarray}
as $T \to \infty$ for some $C_3>0$,
where $\ept$ and $\thr$ are identical to those in Theorem \ref{thm:one}.
\end{prop}
\vspace{2mm}
For proof, see Appendix \ref{app:prop:three}.

We are now ready to prove Theorem \ref{thm:two}.
Proposition \ref{prop:three} implies that the SBS algorithm is consistent in detecting change-points
from wavelet periodograms and cross-periodograms at a single scale $i$,
i.e. all the change-points that are detectable from scale $i$ are identified by applying the SBS algorithm
to the wavelet periodograms and cross-periodograms at the same scale.
Besides, coupled with (B4), the condition on the magnitude of $|W^{(j)}_i(z)|$ in (B1) implies that for each change-point,
the finest scale $i$ at which it is detected satisfies $i \ge \is_T = -\lfloor \alpha\log\log\,T\rfloor$.
Suppose $t=\nu$ is a change-point in $S^{(j)}_i(t/T)$ which can be detected only at the scales coarser than $\is_T$.
Then, the corresponding jump in $\beta^{(j)}_i(t/T)$ is of the magnitude bounded from above by
\begin{eqnarray*}
\left\vert \sum_{i' = -\infty}^{\is_T-1}\left\{S^{(j)}_i\left(\frac{\nu+1}{T}\right) - S^{(j)}_i\left(\frac{\nu}{T}\right)\right\}A_{i, i'} \right\vert
\le C\sum_{i' = -\infty}^{\is_T-1} 2^{i'}A_{i, i'} \to 0,
\end{eqnarray*}
since $A_{i, i'} > 0$ and $\sum_{i' = -\infty}^{-1} 2^{i'}A_{i, i'} = 1$ from \citet{piotr2003},
and thus (B4) is violated.
We conclude that there exists $\is_T$ such that all the change-points $\nu_r, \ r=1, \ldots, N$ are detectable from
examining the scales $i=-1, \ldots, \is_T$.
Since the SBS-MVTS algorithm repeatedly applies the SBS algorithm to the finest $|\is_T|$ scales,
its consistency with the required rates is a simple consequence
of the argument about across-scales post-processing from \citet{cho2012}.

\subsection{Proof of Proposition \ref{prop:one}}
\label{app:prop:one}

Let $t=\lfloor zT \rfloor$. Then we have
\begin{eqnarray*}
c^{(j)}_T(z, \tau)
&=&\E\left\{\sumi\sumk W^{(j)}_i\left(\frac{k}{T}\right)\psi_{i, t-k}\xi^{(j)}_{i, k}
\sum_{i'=-\infty}^{-1}\sum_{k'=-\infty}^\infty W^{(j)}_{i'}\left(\frac{k'}{T}\right)\psi_{i', t+\tau-k'}\xi^{(j)}_{i', k'}\right\}
\\
&=&\sumi\sumk S^{(j)}_i\left(\frac{k}{T}\right)\psi_{i, t-k}\psi_{i, t+\tau-k}.
\end{eqnarray*}
Therefore
\begin{eqnarray*}
&& \frac{1}{T}\sum_{t=0}^{T-1}\left\vert c^{(j)}_T(z, \tau)-c^{(j)}(z, \tau) \right\vert
\le \frac{1}{T} \sum_{t=0}^{T-1} \left\vert\sumi\sumk
\left\{S^{(j)}_i\left(\frac{k}{T}\right)-S^{(j)}_i\left(\frac{t}{T}\right)\right\}
\psi_{i, t-k}\psi_{i, t+\tau-k}\right\vert
\\
&=& \frac{1}{T}\sum_{t=0}^{T-1}\left\vert \sum_{i=-J_T}^{-1} + \sum_{i=-\infty}^{-J_T-1}
\left[ \sumk \left\{S^{(j)}_i\left(\frac{k}{T}\right)-S^{(j)}_i\left(\frac{t}{T}\right)\right\}
\psi_{i, t-k}\psi_{i, t+\tau-k}\right] \right\vert
\equiv I + II,
\end{eqnarray*}
where the cut-off index is set as $J_T=\varrho\log_2T$ for $\varrho\in(0, 1)$.
For all $i=-1, \ldots, -J_T$, the length of support of $\psi_{i, t-k}\psi_{i, t+\tau-k}$
is bounded from above by $K2^{J_T}$ uniformly for some $K>0$.
Therefore, the summands of $I$ are equal to 0 except for those $t$ which are within the distance of $K2^{J_T}$
from any change-point of $S^{(j)}_i(z), i=-1, \ldots, -J_T$.
Then from (B1)--(B2), term $I$ is bounded by
\begin{eqnarray}
&& \frac{NK2^{J_T}}{T}\left\vert \sum_{i=-J_T}^{-1} L^{(j)}_i \sumk\psi_{i, t-k}\psi_{i, t+\tau-k} \right\vert = \frac{NK2^{J_T}}{T}\left\vert \sum_{i=-J_T}^{-1} L^{(j)}_i \Psi_i(\tau) \right\vert
\nonumber \\
&\le& \frac{NK2^{J_T}}{T}\left\vert \sum_{i=-J_T}^{-1} L^{(j)}_i \right\vert
= O\left(\frac{N2^{J_T}\log\,T}{T}\right),
\label{pf:prop:one:one}
\end{eqnarray}
where the first inequality comes from the fact that $\Psi_i(\tau)=O(1)$ uniformly in $\tau$.
The term $II$ is bounded by
$T^{-1}\sum_{t=0}^{T-1} |\sum_{i=-\infty}^{-J_T-1} L^{(j)}_i\Psi_i(\tau)|$ $\le$
$|\sum_{i=-\infty}^{-J_T-1}L^{(j)}_i|$.
Due to the bound imposed on $W^{(j)}_i(z)$, (B1) implies that $|L^{(j)}_i| \le C2^i$
and therefore
$\sum_{i=-\infty}^{-J_T} L^{(j)}_i \le C2^{-J_T} \to 0$,
and combined with (\ref{pf:prop:one:one}) we have $I+II=o(1)$.

As for the relationship between $c_T^{(j, l)}(z, \tau)$ and $c^{(j, l)}(z, \tau)$,
we note (B1)--(B2) implies that the total magnitude of jumps in
$S_i^{(j, l)}(z, \tau)=W_i^{(j)}(z, \tau)W_i^{(l)}(z, \tau)\Sigma_i^{(j, l)}(z, \tau)$ is bounded from above by
$K'\max(L_i^{(j)}, L_i^{(l)}, R_i^{(j, l)})$ for some positive constant $K'$.
Then the proof follows exactly the same arguments as above and thus is omitted.

\subsection{Proof of Proposition \ref{prop:three}}
\label{app:prop:three}

Proposition \ref{prop:three} can be proved by showing that any change-point in $\sigkt$
is detectable from $\tytt^{(k)}$ within the distance of $O(\ept)$.
Let $\sigkt$ have a change-point at $t=\eta_{q_1+q}\equiv \eta$ within a segment $(s, e)$, where $s, \eta_{q_1+q}$ and $e$ satisfy (\ref{lem:cond:one}) and (\ref{lem:cond:two}).
From the compactness of wavelet vector $\bps_i$,
the sequence $\sigktt$ also has a change-point within the interval containing $\eta$,
such that there exists $\bar{\eta} \in [\eta-K 2^{-\is_T}, \eta+K 2^{-\is_T}]$ where
$|\sig^{(k)}_{\bar{\eta}, T} - \sig^{(k)}_{\bar{\eta}+1, T}| > 0$ although it may not be unique.
Let $\bar{\eta}<\eta$. Since such $\bar{\eta}$ still satisfies (\ref{lem:cond:one}) in place of $\eta$,
Proposition \ref{prop:two} implies that
\begin{eqnarray*}
&& \left\{\left|\frac{1}{\bar{\eta}-s+1}\sum_{t=s}^{\bar{\eta}}\sigktt-
\frac{1}{e-\bar{\eta}}\sum_{t=\bar{\eta}+1}^{e}\sigktt\right|
-\left|\frac{1}{\eta-s+1}\sum_{t=s}^\eta\sig^{(k)}\left(\frac{t}{T}\right)-
\frac{1}{e-\eta}\sum_{t=\eta+1}^e\sig^{(k)}\left(\frac{t}{T}\right)\right|\right\}^2
\\
&\le& C\delt^{-2}
\left|\sum_{t=s}^{\bar{\eta}}\left\{\sigktt-\sig^{(k)}\left(\frac{t}{T}\right)\right\}
-\sum_{t=\eta+1}^{e}\left\{\sigktt-\sig^{(k)}\left(\frac{t}{T}\right)\right\}
-\sum_{t=\bar{\eta}+1}^{\eta}\left\{\sigktt+\sig^{(k)}\left(\frac{t}{T}\right)\right\}\right|^2
\\
&\le&
C\delt^{-1}\sum_{t=s}^e|\sigktt-\sig^{(k)}(t/T)|^2 + C'\delt^{-2}\sig^{*2}2^{-2\is_T} \to 0.
\end{eqnarray*}
That is, the CUSUM statistics computed from $\tytt^{(k)}$ are of the same order as those from $\ytt^{(k)}$ around $t=\eta$.
Therefore, the arguments used in Lemmas \ref{lem:one}--\ref{lem:five} also apply to $\tytt^{(k)}$,
and $\heta = \arg\max_{t\in(s, e)}\tbY_{s, t, e}$ satisfies $|\heta-\bar{\eta}| \le c_0\ept$.
Then with $\is_T=-\lfloor\alpha\log\log\,T\rfloor$, we have
$|\heta-\eta| \le |\heta-\bar{\eta}| + |\bar{\eta}-\eta| \le c_0\ept + C\log^{2+\vartheta}\,T$.
Besides, once a change-point is detected within such an interval,
the condition (\ref{uh:cond}) in Step 2.2 does not allow any more change-points to be detected
too close to previously detected change-points, and therefore
any $t \in [\eta-K 2^{-\is_T}, \eta+K 2^{-\is_T}]$ is disregarded from future change-point detection.
Hence despite the bias between $\sigktt$ and $\sigkt$,
the consistency of the SBS algorithm still holds for $\tytt^{(k)}$ in place of $\ytt^{(k)}$.

\bibliographystyle{asa}
\bibliography{fbib}

\end{document}